\documentclass[11pt,a4paper]{article}
\usepackage{jheppub}
    \makeatletter
\def\@fpheader{\relax}
\makeatother

\usepackage{graphicx}
	\graphicspath{ {images/} }
\usepackage{blindtext}	
\usepackage{amsmath,amssymb,amsthm}
\usepackage{latexsym,graphicx,color,subfigure}
\usepackage{enumerate}
\usepackage{color}
\usepackage{hyperref}
\usepackage{slashed}

\newcommand{\D}{\mathcal{D}}
\newcommand{\p}{\partial}
\newcommand{\Tr}{{\rm Tr}}

\newcommand{\bea}{\begin{eqnarray}}
\newcommand{\eea}{\end{eqnarray}}
\def\Tr{ \hbox{\rm Tr}}
\def\tr{ \hbox{\rm tr}}

\def\half{\frac{1}{2}}
\def\a{\alpha}

\def\rh{\rho}
\def\half{\frac{1}{2}}

\def\L{{\rm L}}
\def\R{{\rm R}}
\def\V{{\rm V}}
\def\Y{{\rm Y}}

\def\l{\left}
\def\r{\right}

\title{\begin{flushright}\ \vskip -1.5cm {\small {KEK-TH-2033}}\end{flushright}
\vskip 40pt
{\bf Topological Defects in the Georgi-Machacek Model  }}
\author{Chandrasekhar Chatterjee$^{1, a}$} 
\author{Masafumi Kurachi $^{1, 2, b}$} 
\author{Muneto Nitta$^{1,c}$}
 \affiliation[1]{Department of Physics, and Research and Education Center for Natural Sciences,\\ Keio University,Hiyoshi 4-1-1, Yokohama, Kanagawa 223-8521, Japan}
 \affiliation[2] {Theory Center, High Energy Accelerator Research Organization (KEK), Tsukuba 305-0801, Japan}
 \emailAdd{ chandra(at)phys-h.keio.ac.jp$^{a}$}
 \emailAdd{kurachi@keio.jp $^{b}$}
 \emailAdd{nitta(at)phys-h.keio.ac.jp$^{c}$} 
\date{\today}
\abstract{We study topological defects in the Georgi-Machacek model in a hierarchical symmetry breaking in which 
extra triplets acquire vacuum expectation values 
before the doublet. 
We find a possibility of topologically stable 
non-Abelian domain walls and non-Abelian flux tubes (vortices) in this model.
In the limit of the vanishing $U(1)_\Y$ gauge coupling in which 
the custodial symmetry becomes exact,
the presence of a vortex spontaneously breaks the custodial symmetry,
giving rise to $S^2$ Nambu-Goldstone (NG) modes localized around the vortex 
corresponding to non-Abelian fluxes. 
Vortices are continuously degenerated by these degrees of freedom, 
thereby called non-Abelian. 
By taking into account the $U(1)_\Y$ gauge coupling, 
the custodial symmetry is explicitly broken, 
the NG modes are lifted, and all non-Abelian vortices fall into 
a topologically stable $Z$-string. 
This is in contrast to the SM in which $Z$-strings are non-topological and are unstable in the realistic parameter region.
Non-Abelian domain walls also break the custodial symmetry 
and are accompanied by localized $S^2$ NG modes. 
Finally, we discuss the existence of domain wall solutions bounded by 
flux tubes, where their $S^2$ NG modes match.
The domain walls may quantum mechanically decay 
by creating 
a hole bounded by a flux tube loop, 
and would be cosmologically safe.
Gravitational waves produced from unstable domain walls could be detected by future experiments.}
\begin{document}
\maketitle
\section{Introduction}

The Standard Model (SM) had been established as a reasonable low-energy effective description of the elementary particle physics, and its reliability has further improved after the discovery of the Higgs boson at the LHC \cite{Aad:2012tfa,Chatrchyan:2012xdj}. However, the SM itself has theoretical problems, including hierarchy problem, and it does not explain various things such as the neutrino mass, baryon asymmetry, dark matter, etc. 
These shortcomings of the SM might originate from its Higgs sector. Since the Higgs sector of the SM is constructed in a minimal way to describe the electroweak symmetry breaking, it might be too simple to deal with those problems. This thought motivates us to study extended models of the Higgs sector. Georgi-Machacek (GM) model \cite{Georgi:1985nv,Chanowitz:1985ug} is one of such models, in which a field with $({\bf 3},\bar{{\bf 3}})$ representation of $SU(2)_{\rm L} \times SU(2)_{\rm R}$ global symmetry is introduced in addition to the Higgs doublet. The model incorporates Majorana mass of neutrinos through the type-II see-saw mechanism. An interesting feature of the GM model is that the vacuum expectation value (VEV) of the bi-triplet field can  be arranged in such a way to preserve the diagonal (custodial) $SU(2)$ symmetry of $SU(2)_{\rm L} \times SU(2)_{\rm R}$ at tree level, therefore the magnitude of the VEV of the bi-triplet does not necessarily have to be taken much smaller than the doublet VEV for the consistency with electroweak $\rho$ parameter measurement. It was shown that even at loop level, the custodial symmetry breaking effect coming from the hypercharge interaction is under control \cite{Blasi:2017xmc}. Having additional scalars, including doubly-charged particle, the phenomenology of the GM model is quite rich, and studies for direct detection at hadron colliders \cite{Godunov:2014waa, Chiang:2015amq, Chang:2017niy} and $e^+$-$e^-$ colliders \cite{Chiang:2015rva, Zhang:2017och, Li:2017daq} have been done extensively. The model can be also distinguished from the SM or other extended models by precision Higgs coupling measurement at future experiments \cite{Kanemura:2014bqa}. Various extensions of the GM model has been studied including supersymmetric version \cite{Cort:2013foa}, with fields higher than the triplet representation \cite{Logan:2015xpa}, the one incorporating  an extra singlet to address the dark matter \cite{Campbell:2016zbp}, etc. It was also shown that the strong first order electroweak phase transition, which is necessary for the successful electroweak baryogenesis, could be achieved depending on the parameter choice~\cite{Chiang:2014hia}.

Since the Higgs sector is extended in a non-trivial way, not only the mass spectrum, but also the vacuum structure is quite different from that of the SM. In this paper we discuss  the possible existence of topological defects in the Higgs sector of the GM model.   Topological objects such as monopoles, strings, domain walls may appear when a symmetry group is spontaneously broken and there exist nontrivial  topological numbers or homotopy  groups of the vacuum manifold  of the symmetry breaking. That is, $\pi_n\l(G/H\r) \ne 0$, when symmetry group $G$ is spontaneously broken down to its subgroup $H$. 
As it is well known today that the existence of such  topological objects in early Universe may have cosmological consequences. 
Cosmic strings can be thought of as a reconciliation  between particle physics and cosmology.   In very hot dense early Universe it is assumed that electroweak symmetry or other symmetry is restored. During the process of expansion and cooling down, Universe would have acquired domain structures due to a phase transition.
A variety of topological objects may have been generated due to this phase transition process
due to the Kibble-Zurek mechanism \cite{Kibble:1976sj,Zurek:1985qw}, 
 and  may have disappeared by 
  recombination  after subsequent symmetry breakings or by other dynamical processes.  
The presence of such objects sometimes gives constraints on 
models of elementary particle physics.

The first example of topological vortices in field theory was found in the Abelian-Higgs model \cite{Nielsen:1973cs} 
similar to Abrikosov vortices in a superconductor \cite{Abrikosov:1956sx}. 
Vortices exist whenever the vacuum manifold $G/H$ admits a nontrivial first homotopy group, 
$\pi_1(G/H) \neq 0$. 
There are plenty of other examples of topological objects in Grand unified models such as an $SO(10)$ model where $\mathbb{Z}_2$ vortices can appear \cite{Kibble:1982ae}. 
Vortices behave as cosmic strings in the context of cosmology.
For a review on cosmic strings, see Refs.~\cite{Hindmarsh:1994re, Vilenkin:2000jqa}.  In the context of the SM there exist electroweak strings 
\cite{Nambu:1977ag,Vachaspati:1992fi,Vachaspati:1992jk,Holman:1992rv,James:1992wb,Achucarro:1999it,Eto:2012kb}.
 However, these strings are not topologically stable since the fundamental group of the vacuum manifold 
\begin{eqnarray}
  {SU(2)_{\rm L} \times U(1)_{\rm Y} \over U(1)_{\rm em} } \simeq S^3
  \label{eq:SM-VM}
\end{eqnarray}
  is trivial: $\pi_1(S^3)=0$. When the non-Abelian gauge coupling is turned off, they become so-called semi-local strings which are stable in the type-I superconductor parameter region \cite{Vachaspati:1991dz,Achucarro:1999it}. Among all string solutions, $Z$-strings, 
  containing a flux of $Z_\mu$ particles,  
 can have a parameter region where they become stable \cite{Vachaspati:1992fi, Vachaspati:1992jk,James:1992wb}. 
If $Z$-strings are stable, they were suggested to contribute to electroweak baryogenesis
\cite{Brandenberger:1992ys,Barriola:1994ez}, 
but there is also an objection \cite{Nagasawa:1996gy}.
However, they are unstable in the realistic parameter region of 
the SM. 
Fermion zero modes on $Z$-strings were also discussed in Refs.~\cite{Vachaspati:1992mk,Moreno:1994bk,Earnshaw:1994jj,Garriga:1994wb,Naculich:1995cb,Liu:1995at,Starkman:2000bq,Starkman:2001tc,Graham:2011fw},
in which it was argued that these zero modes may destabilize $Z$-strings.
Moreover, 
endpoints of strings are attached by a monopole or an anti-monopole. 
Therefore, the $Z$-strings can quantum mechanically decay by 
a nucleation of a monopole-anti-monopole pair and 
are therefore at most
metastable in this sense even in the stable parameter region \cite{Preskill:1992ck}. 
$Z$-strings ending on monopoles were suggested to generate 
primordial magnetic fields in cosmology 
  \cite{Vachaspati:2001nb,Poltis:2010yu}.
Saddle point solutions corresponding to monopole and anti-monopole connected by a $Z$-string  
are known as sphalerons \cite{Klinkhamer:1984di}.

    Other than monopoles and strings, there can exist domain walls or kinks when a discrete symmetry is spontaneously broken. 
Domain wall or kink configurations depend on one spatial direction,  
appearing as a partition between two different vacua during phase transitions. 
Particulary common kinks appearing in various physical systems are sine-Gordon kinks 
\cite{Perring:1962vs,Manton:2004tk, Rajaraman:1982is} discussed for long time starting from condensed matter physics, such as  Josephson junctions of two superconductors \cite{Ustinov} to cosmology \cite{Vachaspati:2006zz}.
     Stable domain wall solutions are cosmologically forbidden, and so any model with stable domain wall solutions is ruled out. However,  we can have  domain walls which are separated from vacuum by finite energy barrier. In this case, domain walls can decay by nucleation of a hole, typically bounded by a closed string  \cite{Kibble:1976sj, Kibble:1982dd, Vilenkin:1982ks, Everett:1982nm, Zeldovich:1974uw, Vilenkin:2000jqa}. 
     For instance, axion models have a cosmological domain wall problem when an axion string is attached by multiple domain walls, while they are free from that problem when it is attached by one domain wall 
     \cite{Kawasaki:2013ae}.

       In this paper, we find out that there exist similar nontrivial topological structures (domain walls and electroweak strings) in the GM model if we consider a hierarchical symmetry breaking of the symmetry group $G$ in two stages, namely only the triplets obtain the VEV first, then the doublet obtain its VEV later. 
       The opposite ordering allows only the same vacuum manifold with that of the SM in Eq.~(\ref{eq:SM-VM}) and is not new.
       This kind of hierarchy in symmetry breaking scales may have occurred during expansion and cooling periods of the early Universe.  Similarly to the SM we have an electroweak gauge symmetry group $SU(2)_\L \times U(1)_\Y$ which we denote by $G_\Y$. The global symmetry group of the potential is found to be a larger group $SU(2)_\L \times SU(2)_\R$ as the SM, which is the same as the symmetry group of the Lagrangian if we ignore the $U(1)_\Y$ gauge interaction. We denote this enlarged group by $G$ and then  
 we have the full symmetry breaking  of $G$ as
\begin{eqnarray}
\label{fullbreaking1}
G = SU(2)_\L \times SU(2)_\R \stackrel{\Phi_v}{\longrightarrow} H_3 = \mathbb{Z}_2 \times  SU(2)_V \stackrel{\Psi_v}{\longrightarrow} H_2 = SU(2)_\V 
\end{eqnarray}
	where $\Phi_v$ and $\Psi_v$ are triplet and doublet fields which acquire nonzero VEVs during  each symmetry breaking stages. Here $SU(2)_\V $ is the diagonal subgroup of $G$, known as the custodial symmetry which remains unbroken throughout.
 Then we consider the case when $U(1)_\Y [\subset SU(2)_\R]$ is gauged. In this case, $SU(2)_\R$ is explicitly broken by the gauge field interaction. We find the symmetry breaking structure as
\begin{eqnarray}
\label{fullbreaking2}
G_\Y = SU(2)_\L \times U(1)_\Y \stackrel{\Phi_v}{\longrightarrow} H^\Y_3 = \mathbb{Z}_2 \times  U(1)_{\rm em} \stackrel{\Psi_v}{\longrightarrow} H^\Y_2 = U(1)_{\rm em} .
\end{eqnarray}
 Here $U(1)_{\rm em}$ is the electromagnetic gauge group. For both the above cases  we find  nontrivial homotopy groups as
 \begin{eqnarray} 
  \pi_0(H_3 /H_2) =  \mathbb{Z}_2, \quad \pi_1(G/H_3) = \mathbb{Z}_2,
 \end{eqnarray}
 implying the existence of domain walls and vortices, respectively at each stage. 
 
 More precisely, if we consider the simplest case of the symmetry breaking of $H_3 /H_2$, we have a domain wall solution due to $\pi_0(H_3 /H_2) =  \mathbb{Z}_2$. We find that this system has stable non-Abelian sine-Gordon kink solutions \cite{Nitta:2014rxa, Eto:2015uqa, Nitta:2015mma,Nitta:2015mxa}.  These solutions spontaneously break the custodial $SU(2)_\V$ symmetry to a $U(1)$ subgroup, generating Nambu-Goldstone (NG) modes known as orientational zero modes 
 \cite{Ritz:2004mp,Nitta:2012wi,Nitta:2012rq}. 
 These are collective coordinates giving an orientation of the unbroken $U(1)$ group within $SU(2)_\V$ on the coset $S^2 \simeq SU(2)_\V/ U(1)$.   Stable domain wall solutions are cosmologically forbidden, and so in this sense the GM model (with the hierarchical symmetry breaking) could have been ruled out. However, this is not the case because of $\pi_1(G/H_3) = \mathbb{Z}_2$ supporting  $ \mathbb{Z}_2$ vortex solutions. It is known that the existence of vortices which bound a domain wall can make domain walls to decay \cite{Preskill:1992ck} as axion domain walls, as mentioned above.

  The $\mathbb{Z}_2$-string was originally discussed by Nielsen and Olesen \cite{Nielsen:1973cs} in an $SU(2)$ gauge theory coupled with two adjoint scalars. The $\mathbb{Z}_2$-strings and more generally $\mathbb{Z}_N$-strings in non-Abelian gauge theories were discussed for instance in Refs.~\cite{Olive:1982zh, Hindmarsh:1985xc, Markov:2004mj, Kneipp:2003ue}. In the GM model, the string solution is similar to a $\mathbb{Z}_2$-string discussed there.  
However, the $\mathbb{Z}_2$-string that we construct here is found to  be non-Abelian in the sense that it contains non-Abelian flux directed
  along generic direction inside an internal space in the absence of the $U(1)_\Y$ gauging.\footnote{ A non-Abelian $\mathbb{Z}_2$-string  with non-Abelian moduli  was discussed in the case of $\mathcal{N} = 1^*$ supersymmetric gauge theories in Ref~\cite{Markov:2004mj}.} In the limit of vanishing  $U(1)_\Y$ gauge coupling of the hypercharge,  these vortices spontaneously break the  custodial $SU(2)_\V$ symmetry to a $U(1)$ subgroup inside the vortex, 
  This generates a continuous degeneracy in the whole vortex solutions, 
  described by NG modes  living on the $S^2 \simeq SU(2)_\V/U(1)$ 
  as the same as the case of non-Abelian domain walls mentioned above. So the flux can be directed along any generic direction on  $S^2 \simeq SU(2)_\V/U(1)$. This kind of NG modes are known as orientational moduli of a non-Abelian vortex.

  Non-Abelian vortices and their non-Abelian orientational moduli  have been investigated extensively in the literature in great details in supersymmetric gauge theories \cite{Hanany:2003hp, Auzzi:2003fs, Auzzi:2003em, Hanany:2004ea, Shifman:2004dr, Gorsky:2004ad, Eto:2005yh, Eto:2006cx, Eto:2006db,Eto:2006pg} and color-flavor locked phase in dense QCD \cite{Balachandran:2005ev,Nakano:2007dr,Nakano:2008dc, Eto:2009kg,Eto:2009bh,Eto:2009tr, Hirono:2010gq,Vinci:2012mc, Cipriani:2012hr,Eto:2013hoa, Kobayashi:2013axa,Chatterjee:2015lbf,Chatterjee:2016ykq,Chatterjee:2016tml}. They may play crucial role in understanding the confinement mechanism and duality in non-Abelian gauge theories.  The dual confinement which is known as monopole-vortex complex where magnetic monopoles are confined by the attachment of flux tubes in hierarchical symmetry breaking are discussed in Refs.~\cite{Auzzi:2003em,Hanany:2004ea,Shifman:2004dr, Eto:2006pg, Eto:2006dx, Nitta:2010nd, Konishi:2012kx, Cipriani:2011xp, Chatterjee:2014rqa, Chatterjee:2009nx, Chatterjee:2009pi}.  The dual confinement in dense QCD was discussed in Ref.~\cite{Eto:2011mk}.
 
 In our case, a non-Abelian vortex is attached by a non-Abelian domain wall where 
 the both $S^2$ moduli match, in the presence of the VEVs of the both triplets and dublet.
 In the limit of the vanishing interaction between the triplets and doublet, these $S^2$ modes can 
 be different, and in this case, actually the domain wall disappear and the total configuration 
 is reduced to a global vortex. 
 
  In the presence of $U(1)_\Y$ gauging, the custodial $SU(2)_\V$ symmetry is explicitly broken, 
  and consequently the degeneracy of vortex solutions on full $S^2$ is lost and is reduced 
  to the north and south poles and the equator circle. 
The same mechanism was studied in dense QCD \cite{Vinci:2012mc,Cipriani:2012hr} and supersymmetric QCD \cite{Konishi:2012eq}.
  In this case we find that there exist two kinds of string solutions. First is a topologically stable $Z$-string corresponding to the poles on $S^2$ for which $SU(2)_\L$ and $U(1)_\Y$ gauge fields are parallel.
	 We should emphasize here that our ${Z}$-strings are topologically stable, in contrast to $Z$-strings in the SM.\footnote{There exist topologically stable global electroweak strings in the two-Higgs doublet model \cite{Dvali:1993sg}.
	 } 
 Other than the $Z$-strings, we also find $W$-strings in which the $SU(2)_\L$ gauge fields are orthogonal to the the $U(1)_\Y$ gauge field and naturally the flux of the $W$-strings consist of $W$ bosons and no contribution from $U(1)_\Y$. 
 The $W$-strings have an $S^1$ degeneracy and live on the equator circle of the $S^2 \simeq SU(2)_\V/U(1)$. 
 The $W$-strings have higher energy than the $Z$-strings and are unstable 
 to decay to a $Z$-string.

This paper is organized as follows.  In Sec.~\ref{GMsymmetry} we briefly introduce the GM model first and then discuss full two-stage symmetry breakings in two subsequent subsections. In Sec.~\ref{DWconstruction} we derive domain wall solution and discuss its orientational zero modes. In Sec.~\ref{vortices} when the only triplets develop VEVs, 
we find non-Abelian vortex solutions in the limit of the vanishing $U(1)_{\rm Y}$ gauge coupling by constructing profile functions numerically. We also discuss orientational zero modes. We then switch on $U(1)_{\rm Y}$ gauge coupling and obtain $Z$-strings and $W$-strings.
In Sec.~\ref{DVcomplex} we discuss the most general solutions in the presence of VEVs of the doublet and triplets.
  In the decoupling limit of the doublet and triplets, the flux tube of the triplet is accompanied with a winding of the doublet component and becomes a global vortex. In the presence of interaction between the doublet and triplets, 
  it becomes a domain wall-vortex composite state where a domain wall is bounded by a vortex flux tube.
We then  discuss the quantum mechanical decay of a domain wall by creating a hole bounded by a vortex loop and estimate the decay rate. Sec.~\ref{summary} is devoted to a summary and discussion.

\section{Symmetry  of  the Georgi-Machacek Model}\label{GMsymmetry}
Let us start by reviewing the GM model first then we will discuss the hierarchical symmetry breaking in this section.
\subsection{The Georgi-Machacek model}
\label{GMmodel}
The scalar sector of the SM supports an $G = SU(2)_\L \times SU(2)_\R$ accidental symmetry. Out of the full symmetry group $G$, the $SU(2)_\L$ and $U(1)_{\rm Y}$ inside $SU(2)_{\rm R}$ is gauged to produce mass of the weak gauge bosons.
It  can be shown   that the diagonal generator of $SU(2)_\R$ can be defined as  the hypercharge. In this case the doublet scalar $\psi^T = (\psi_1, \psi_2)$ is written in $( \bar 2, 2)$ form as
$\Psi = \left(
\begin{array}{ccc}
\psi^*_2  &  \psi_1 \\
- \psi^*_1  & \psi_2
\end{array}
\right)$and it breaks the symmetry group $G$ generating mass to $W^\pm_\mu$.
In this process, the VEV of the doublet keeps the diagonal subgroup $SU(2)_\V$ unbroken which is known as the custodial symmetry. This model can be generalized by adding more scalars in different representations
by keeping the symmetry structure the same, at least locally.  
The GM model \cite{Georgi:1985nv} is one of such extensions. In this case extension can be done by using an extra complex triplet $\Phi^T = (\phi_{++}, \phi_{+}, \phi_{0})$ which can be written in $( \bar 3, 3)$ form with the help of  another real triplet scalar $\zeta^T = (\zeta_{+}, \zeta_{0}, \zeta_{-})$ as
\begin{eqnarray}
\Phi(x) = (\phi_c, \zeta, \phi ) =  \left(
\begin{array}{ccccc}
\,\,\,\phi_{0}^*  &  &  \zeta_+ & & \phi_{++}  \\
- \phi_{+}^*  &  & \zeta_0  & & \phi_{+} \\
\,\,\,\,\phi_{++}^*  & &  \zeta_- &  & \phi_{0}
\end{array}
\right), 
\end{eqnarray}
where $ \phi _c = C_3 \phi^*$.
\footnote{
The notation of the matrices is as follows:
\tiny\begin{eqnarray}
 T^1 = \frac{1}{\sqrt 2}
\left(
\begin{array}{ccc}
0  &  1 &  0 \\
1  & 0  & 1  \\
0  & 1  & 0  
\end{array}
\right), T^2 = \frac{1}{\sqrt 2}
\left(
\begin{array}{ccc}
0  &  -i &  0 \\
i  & 0  & -i  \\
0  & i  & 0  
\end{array}
\right), T^3 = 
\left(
\begin{array}{ccc}
1  &  0 &  0 \\
0  & 0  & 0  \\
0  & 0  & -1  
\end{array}
\right), 
\Tr T^aT^b = 2 \delta^{ab}, 
C_3 =  \left(
\begin{array}{ccccc}
0  &  &  0 &  & 1 \\
0  &  & -1  &  & 0  \\
1  & &  0  & & 0  
\end{array}
\right), \tau^a = \half \sigma^a.
\end{eqnarray}
\label{generators}}

 We start with the Lagrangian density as
 \begin{eqnarray}
 \label{L1}
\mathcal{L} = - \frac{1}{8}\Tr W_{\mu\nu}^2 - \frac{1}{4} B_{\mu\nu}^2 + \half \Tr(\D_\mu\Phi)^\dagger\D_\mu\Phi + \half \Tr(\D_\mu\Psi)^\dagger\D_\mu\Psi - V(\Phi,\Psi), 
\end{eqnarray}
where $W_{\mu\nu} = \p_\mu W_\nu - \p_\mu W_\mu - i g_{\rm W} [W_\mu, W_\nu] $ , $B_{\mu\nu} = \p_\mu B_\nu - \p_\nu B_\mu$  are the field strengths of the gauge fields of $SU(2)_\L$  and $U(1)_\Y$
 gauge symmetry, respectively, and the covariant derivatives are defined by
\begin{eqnarray}
\label{Dmu}
\D_\mu \Phi = (\p_\mu- i g_{\rm W} W_\mu^aT^a )\Phi + i g_{\rm Y} \Phi B_\mu T^3, 
\quad
\D_\mu \Psi = (\p_\mu- i g_{\rm W} W_\mu^a\tau^a )\Psi + i g_{\rm Y} \Psi B_\mu \tau^3.
\end{eqnarray}
Here $g_{\rm W}$ and $g_{\rm Y}$ are the coupling constants of $SU(2)_\L$ and $U(1)_\Y$ gauge interactions, respectively 
and $T^a$ and $\tau^a$ are the  triplet and doublet representations, respectively, of the generators of the $SU(2)$ algebra.$^{\ref{generators}}$
The potential that serves our purpose can be expressed as 
\begin{eqnarray}
V(\Phi,\Psi) &= & \lambda_1\l( \Tr \Phi^\dagger\Phi - 3v_3^2\r)^2 + \lambda_2\l[3 \Tr \Phi^\dagger\Phi \Phi^\dagger\Phi - \Tr \l( \Phi^\dagger\Phi\r)^2\r] + \lambda_3\l( \Tr \Psi^\dagger\Psi - v_2^2\r)^2 \nonumber\\
&+& \lambda_4\l( \Tr \Psi^\dagger\Psi \Tr \Phi^\dagger\Phi - 2 \Tr( \Psi^\dagger \tau^a \Psi \tau^b)  \Tr (\Phi^\dagger T^a \Phi T^b)\r).
           \label{potential}
\end{eqnarray}
Here we consider the parameter region $\lambda_1 + \lambda_3 > 0, \lambda_2 > 0, \lambda_4 > 0.$
We have written minimum number of terms in the potential required to fulfill our purposes 
\cite{Chanowitz:1985ug, Logan:2015xpa}.   

Now let us
discuss symmetry of the Lagrangian defined in Eq.~(\ref{L1}).  We first discuss symmetry of the potential which is the same as that of the case in which the gauge coupling of $U(1)_\Y$ (hypercharge)
is turned off, {\it i.e.} $g_{\rm Y} = 0$. Later we discuss the effect of $U(1)_\Y$ gauging.
The potential in Eq.~(\ref{potential}) is invariant under an enlarged symmetry group
\begin{eqnarray}
\label{sg-Z2}
   G_0 =  {SU(2)_\L \times SU(2)_\R \over ({\mathbb Z}_2)_{\rm V}}.
\end{eqnarray}
To understand the action of $G_0$ over the fields, let us define any element in pair as $g = (g_\L, g_\R)$ 
in the universal covering group 
\begin{eqnarray}
\label{sg}
   G =  SU(2)_\L \times SU(2)_\R .
\end{eqnarray}
The action of the group elements on the triplets and doublet can be defined as
\begin{eqnarray}
\label{phi'}
 && \Phi'(x) = g_\L(T^a)\,\Phi (x)\, g_\R^\dagger(T^a), \quad g_{\L/\R}(T^a) = e^{i\a_{\L/\R}^a T^a },  \\
 && \Psi'(x) = g_\L(\tau^a)\,\Psi (x)\, g_\R^\dagger(\tau^a), \quad g_{\L/\R}(\tau^a) = e^{i\a_{\L/\R}^a \tau^a },
\end{eqnarray}
respectively. 
Then, 
$(\mathbb{Z}_2)_{\rm V}$ in the denominator in Eq.~(\ref{sg-Z2}) is given by
$(\mathbb{Z}_2)_{\rm V} = \l\{ (1, 1), (-1, -1)\r\}$, since this group does not act on these fields. 
The full center $\mathbb{Z}_2 \times \mathbb{Z}_2$ 
of the symmetry group $G$ is fully unseen by the triplet field $\Phi$,
while 
$(\mathbb{Z}_2)_{\rm A} = \l\{ (1, 1), (1, -1)\r\}$ acts on the doublet field $\Psi$ 
although it does not act on the triplet fields $\Phi$ 
(and it is spontaneously broken when the doublet aquires a VEV). \footnote{ 
The center of the symmetry group $G$  can be written  as the Klein-4 group $V_4 = \mathbb{Z}_2 \times \mathbb{Z}_2 $. Elements of  the center can be expressed in pairs as
$ V_4 = \l\{e = (1, 1), a = (-1, -1), b = (1, -1), c = (-1, 1)\r\}$.
This group have three normal subgroups and can be written as
$M_v = \{e, a\}, \quad M_1 = \{e, b\}, \quad M_2= \{e, c\}$.
Any two of them are {\it permutable complements} to each other. {\it So $V_4$ can also be written as internal direct product of
any two of the above subgroups}. One of them, namely $M_v$, is the center of $SU(2)_\V$, the diagonal subgroup of  $G$.
\label{center}
}
Hereafter, 
we work with the universal covering group. 
\subsection{
Symmetry breaking}
In this paper our purpose  is to introduce a hierarchical symmetry breaking of the full symmetry group $G$. So our intension is to break the symmetry in two stages, first by the triplet field $\Phi(x)$
and then by the doublet field $\Psi(x)$.\footnote{The VEVs of the fields are temperature dependent in reality and the mass term in the potential can be expressed as $\scriptstyle{C_1\l[\l(\frac{T}{T_{c_1}}\r)^2 - 1 \r]\Tr\l(\Phi^\dagger \Phi\r) + C_2 \l[\l(\frac{T}{T_{c_2}}\r)^2 - 1 \r]\Tr\l(\Psi^\dagger \Psi\r) }$. As  the universe cools down, 
the temperature ($T$) may have reached a value $T_{c_1} > T>T_{c_2}$ where only first transition could occur. 
} 
The details of the symmetry breaking process, temperature dependence and fine-tuning of parameters can be discussed elsewhere. In this paper we just assume the possibility of 
two stage symmetry breaking
and for this purpose we assume $v_3 > v_2$. For technical  reason we keep $v_3 >> v_2$, however, for practical purposes this constraint may not be very strict.

Let us now understand the symmetry breaking in details. $SU(2)_\L $ and $ SU(2)_\R$  groups act on the triplet field $\Phi$ from left and right accordingly as described in Eq.~(\ref{phi'}). Now we introduce the triplet VEV as 
\begin{eqnarray}
\label{phiv}
\Phi_v =  v_3\left(
\begin{array}{ccccc}
1  &  &  0 & & 0 \\
0  &  & 1  & & 0  \\
0  & & 0  &  & 1
\end{array}
\right).
\end{eqnarray}
As it can be understood easily from the Eq.~(\ref{phi'}) that
the diagonal group elements $(g, g)$ of $SU(2)_\L $ and $ SU(2)_\R$ does not act on the VEV.  
So $\Phi_v$ breaks the symmetry group  $SU(2)_\L \times SU(2)_\R$  and
 keeps the diagonal subgroup $SU(2)_\V$ unbroken.
 Including a discrete group, the unbroken group $H_3$ inside the universal covering group $G$ is found to be
 \begin{eqnarray}
 \label{h3}
H_3  
=  SU(2)_\V  \times (\mathbb{Z}_2)_{\rm A}, \quad 
  (\mathbb{Z}_2)_{\rm A}  = \l\{ (1, 1), (1, -1)\r\}.
\end{eqnarray} 
The $ \mathbb{Z}_2$ in Eq.~(\ref{h3}) is one of the normal subgroups of the center.$^{\ref{center}}$
The  existence of this  $ \mathbb{Z}_2$ in $H_3$ can be understood easily if we consider action of   elements on the doublets and triplets separately.

 The vacuum manifold is found to be
\begin{eqnarray}
\frac{G}{H_3} = \frac{SU(2)_\L \times SU(2)_\R}{(\mathbb{Z}_2)_{\rm A} \times  SU(2)_\V} \simeq \frac{SU(2)}{\mathbb{Z}_2} 
\simeq SO(3) \simeq {\mathbb R}P^3. 
\end{eqnarray}
Since $G$ is simply connected we may express the fundamental group as
\begin{eqnarray}
\pi_1\l(\frac{G}{H_3}\r) \simeq \pi_0(H_3) = \mathbb{Z}_2,
\end{eqnarray}
implying the existence of a ${\mathbb Z}_2$ string. 

The $\mathbb{Z}_2$ in $H_3$ in Eq.~(\ref{h3}) 
nontrivially acts on  the doublet,  and so it is broken when the doublet acuires a VEV 
during the second symmetry breaking. 
The  invariance of the potential under  the group $H_3$ can be understood clearly
once we insert the value of $\Phi_v$ into the potential.  
After setting the triplet field $\Phi$ in its vacuum value $\Phi = v_3 \mathbf{1}_{3\times 3}$  the potential for $\Psi$ field is found to be\,
 \begin{eqnarray}
 \label{psipot}
V\l(v_3 \mathbf{1}_{3\times 3},\Psi\r) 
=  \lambda_3\l( \Tr \Psi^\dagger\Psi - v_2^2\r)^2 
+ 2  \lambda_4 v_3^2 \l(2 \Tr \Psi^\dagger\Psi  -  |\Tr\Psi |^2  \r). 
\end{eqnarray}
We then find that the doublet field takes the form 
\begin{eqnarray}
\Psi_v = \pm   \frac{v_2}{\sqrt 2}
\left(
\begin{array}{ccc}
1  &   0 \\
  0 &  1 
\end{array}
\right)
\label{psiv}
\end{eqnarray}
in the vacua.
This confirms the existence of $\mathbb{Z}_2$ and its breaking.  
\subsection{The effect of $U(1)_\Y$ symmetry}\label{Ysymmetry1}
So far we discussed the symmetry of the potential and its breaking. However, when we introduce the  $U(1)_\Y \in SU(2)_\R$ as local symmetry, 
the structure of the symmetry breaking changes a little. In this case $SU(2)_\R$ is explicitly broken and we may write the full symmetry group $G_\Y$ of the Lagrangian
as 
\begin{eqnarray}
G_\Y =  SU(2)_\L \times U(1)_\Y.
\end{eqnarray}
The VEV of the triplet fields  $\Phi_v$ in Eq.~(\ref{phiv}) breaks  $G_\Y$ to $H^\Y_3 = \mathbb{Z}_2\times U(1)_{\rm em}$. Here $U(1)_{\rm em}$ is the 
gauge group of electromagnetic theory and is defined as a subgroup of the custodial symmetry group $SU(2)_\V$. The  VEV of the
doublet $\Psi_v$ in Eq.~(\ref{psiv}) breaks $H^\Y_3$ to $U(1)_{\rm em}$. The full symmetry breaking in two stages is expressed in Eq.~(\ref{fullbreaking2}).
In this case the vacuum manifold of the first symmetry breaking is different from what we found previously, however the fundamental group remains the same 
\begin{eqnarray}
\frac{G_\Y}{H_3^\Y} = \frac{ SU(2)_\L \times U(1)_\Y}{ \mathbb{Z}_2 \times  U(1)_{\rm em}}, \qquad \pi_1\l(\frac{G_\Y}{H^\Y_3}\r) = \mathbb{Z}_2.
\end{eqnarray}
We shall see that  vortices and domain walls can be constructed  in this case also. Only difference is the existence of electromagnetism and consequences  would be  discussed  later. 

The kinetic term of the scalar field is given as
$\D_\mu \Phi = (\p_\mu- i g_{\rm W} W_\mu^aT^a )\Phi + i g_{\rm Y} \Phi B_\mu T^3$. 
In this case we define the well known  $Z_\mu$ boson and $A_\mu$ electromagnetic gauge field as
\begin{eqnarray}
{A}_\mu  =\,\, \sin\theta_{\rm W} W_\mu^3  +  \cos\theta_{\rm W}   B_\mu,\quad
{Z}_\mu  =\,\, \cos\theta_{\rm W} W_\mu^3 -  \sin\theta_{\rm W}   B_\mu
\end{eqnarray}
where $ \cos\theta_{\rm W} = \frac{g_{\rm W}}{\sqrt{g_{\rm W}^2+g_{\rm Y}^2}}$ and $ \sin\theta_{\rm W} = \frac{g_{\rm Y}}{\sqrt{g_{\rm W}^2+g_{\rm Y}^2}}$. Using this Eq.~(\ref{Dmu}) can be expressed as
\begin{eqnarray}
\D_\mu \Phi &=& \l(\p_\mu - i g_{\rm W} \sum_{\pm} W_\mu^{\pm}T^\pm\r) \Phi  - i Z_\mu \l(g_{\rm W}  \cos\theta_{\rm W}    T^3 \Phi  +  g_{\rm Y} \sin\theta_{\rm W}  \Phi  T^3\r) \nonumber\\
&& \phantom{xxxxxxxxxxxxxxxxxxxxxxx}- i g_{\rm W} \sin\theta_{\rm W} A_\mu\l(T^3  \Phi  -  \Phi T^3  \r) .
\end{eqnarray}
So naturally we can define electric charge $e = \frac{g_{\rm W}g_{\rm Y}}{\sqrt{g_{\rm W}^2+g_{\rm Y}^2}}$. At the vacuum when first symmetry breaking occurs  $\Phi = v_3 \mathbf 1$, the $A_\mu$ interaction vanishes.
Similar situation occurs for the doublet also. After full symmetry breaking
the masses of gauge fields are given by 
\begin{eqnarray}
m_Z^2 = \l(2 v_3^2 + \frac{1}{4} v_2^2\r)g_{\rm Z}^2 , \qquad m_{W}^2 = \l(2 v_3^2 + \frac{1}{4} v_2^2\r)g_{\rm W}^2,\qquad  g_{\rm Z} = \sqrt{g_{\rm W}^2 + g_{\rm Y}^2}. 
\end{eqnarray}
We are assuming $v_3 > v_2$ case so the masses are dominated by the VEV  $v_3$ of the triplets.

\section{Non-Abelian Domain Walls }
\label{DWconstruction}
Domain wall solution occurs whenever the discrete symmetries of  a field theory are spontaneously broken in the ground state.  In the case of the situation
described above we saw that  after the first phase transition which is triggered by the triplet VEV, our potential is invariant under a symmetry group $H_3$ which contains a discrete subgroup of the original symmetry group $SU(2)_\L \times SU(2)_\R$ and described as
$\mathbb{Z}_2 \times  SU(2)_\V$. Now if we observe  the second phase transition as in Eq.~(\ref{fullbreaking1}),
we may notice that it breaks $\mathbb{Z}_2$, 
implying the existence of a domain wall.

Let us first start with $g_{\rm Y} =0$, that is, without $B_\mu$ interaction. Since in this case all the existing gauge fields become massive after the first phase transition, we may ignore them while constructing  
domain walls. Hence, let us consider the following reduced Lagrangian constructed by inserting the vacuum expectation value of the triplet as  $\Phi = v_3 {\mathbf 1}_{3\times 3}$ into Eqs.~(\ref{L1}) and (\ref{potential}) and also by setting all the gauge fields equal to zero:
 \begin{eqnarray}
 \label{L2}
\mathcal{L}_2 =  \half \Tr(\p_\mu\Psi)^\dagger\p_\mu\Psi - \lambda_3\l( \Tr \Psi^\dagger\Psi - v_2^2\r)^2 - 2  \lambda_4 v_3^2 \l(2 \Tr \Psi^\dagger\Psi - \Tr\Psi^\dagger \Tr\Psi  \r).
\end{eqnarray}
Now we  define our domain wall static ansatz along as 
\begin{eqnarray}
\label{DW1}
\Psi(x)_{\rm dw} &=&  \frac{v_2}{\sqrt 2} \psi(x) \exp\l[i\phi(x)\,\tau^3\r].
\end{eqnarray}
Here the fields $\psi(x)$ and $\phi(x)$ are functions of a single spacial coordinate  assuming that  the center of the wall will be on the orthogonal plane at $x =0$. 
After inserting the ansatz into the Lagrangian we find
\begin{eqnarray}
- \frac{ \mathcal{L}_2}{ v_2^2}  = \half (\p_i\psi)^2 + \frac{1}{8}\psi^2 (\p_i \phi)^2 + \lambda_3 v_2^2 \l(  \psi^2 - 1\r)^2 + 2  \lambda_4 v_3^2  \psi^2 \l(1 - \cos\phi \r).
\end{eqnarray}
Now, to make the point clearer, let us consider the extreme situation when $\frac{\lambda_3}{\lambda_4} \gg \frac{v_3^2}{v_2^2}$. In this case we may  assume $\psi(x)$ can be set at one of the vacua,
say $\psi(x) = 1$, then we may have domain wall  in $\phi$
with the boundary conditions
 \begin{eqnarray}
\phi(x = \infty) = 2\pi, \qquad  \phi(x = -\infty) = 0, 
\end{eqnarray}
and this corresponding to $\mathbb{Z}_2$ transformation at the boundary as $\Psi(- \infty)\stackrel{\phi}{ \longrightarrow} - \Psi(\infty)$. In this case the 
Lagrangian reduces to 
\begin{eqnarray}
\label{DWH}
- \frac{ \mathcal{L}_\phi}{ v_2^2} =  \frac{1}{8}\l[ (\p_i \phi)^2  + 8 \mu^2     \l(1 - \cos\phi \r)\r], \qquad\mu^2 =  2\lambda_4 v_3^2. 
\end{eqnarray}
This is identical to the sine-Gordon  model and a domain wall solution interpolating between the two vacua can be written as
\begin{eqnarray}
\phi(x) =  4 \tan^{-1} e^{\pm 2\mu x},
\end{eqnarray}
where 
the width of the domain wall is given by $\delta_{\rm dw} \sim  \mu^{-1}$ and   the energy per unit area can be written as
\begin{eqnarray}
T_{\rm dw} = 4\mu v_2^2 . 
\end{eqnarray}
\begin{figure}[!htb]
\centering
\includegraphics[totalheight=4cm]{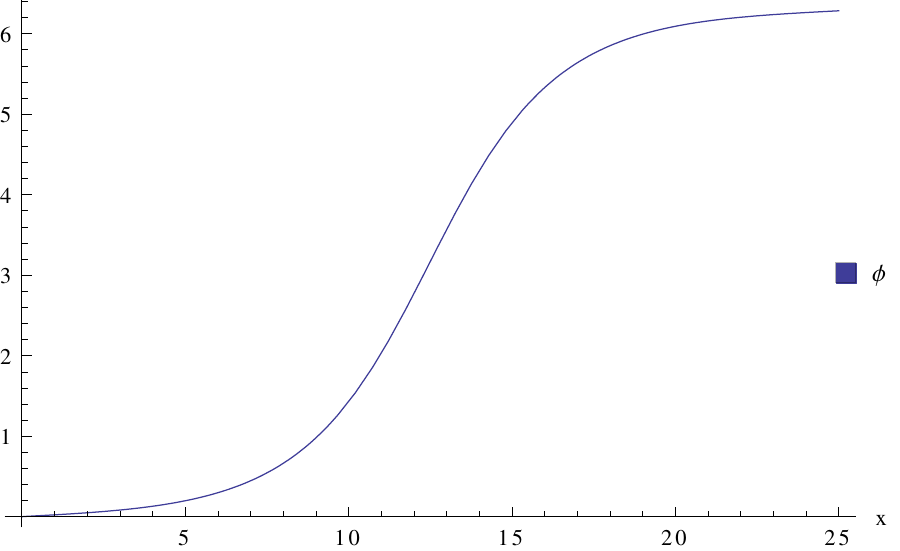}
\caption{A plot of Sine-Gordon kink}
\label{sinegordon}
\end{figure}
The shape of the solution is shown Fig. \ref{sinegordon}.

Here we may say few wards on the solution ansatz in Eq.~(\ref{DW1}). 
As it is discussed before that the vacuum after the second symmetry breaking preserves the $SU(2)_\V$ custodial symmetry. The presence of 
a domain wall configuration
spontaneously breaks the $SU(2)_\V$ custodial symmetry into a $U(1)$ subgroup 
in the vicinity of the wall . 
 It can be checked easily that  at boundary where $\phi = (0, 2\pi)$ the custodial symmetry is recovered. 
This spontaneous breaking of the custodial symmetry  generates NG modes.  These are `orientational' zero modes on the domain wall surface
parameterizing the coset space $S^2 \simeq SU(2)_\V/ U(1)$. The existence of these modes allow us to define the ansatz in generic direction on $S^2$ by a global transformation as
\begin{eqnarray}
\Psi(\xi^\a, x) &=&  G(\xi^\a) \frac{v_2}{\sqrt 2} \psi \exp\l[i\phi\tau^3\r] G^\dagger(\xi^\a) = \frac{v_2}{\sqrt 2} \psi(x) \exp\l[i\phi(x)\,\mathbf{\hat n}\r], \\
\text{where} \, && \mathbf{\hat n} = G(\xi^\a) \tau^3 G^\dagger(\xi^\a),\quad
G(\xi^\a) = 
\left(
\begin{array}{ccc}
\cos\frac{\xi^1}{2}   &   -   \sin \frac{\xi^1}{2}e^{- i\xi^2}  \\
 \sin \frac{\xi^1}{2} e^{i\xi^2}&   \cos\frac{\xi^1}{2} 
\end{array}
\right), 
\end{eqnarray}
where 
$\xi^\a$ are the coordinate angles defined on $S^2$ and 
$\tr \,(\mathbf{\hat n}^2)=1$.
The effective theory of the NG modes should be an $O(3)$ sigma model on the $2+1$ dimensional world-volume.

Now we should talk about the case when $g_{\rm Y}\ne 0$. Since domain wall construction depends only on scalar fields but not on gauge field interactions, and so the introduction of 
$U(1)_\Y$ would not effect the construction at the tree level. 
The radiative correction will break $SU(2)$ custodial symmetry explicitly, and hence $S^2$ moduli will be lifted.
Also,
the presence of electromagnetic field would generate interaction with $S^2$ zero modes. So the effective action of static domain wall would be described by an $SO(2)$  gauged $O(3)$ sigma model
living in $2+1$ dimensional hyperplane interacting with electromagnetic gauge field living in $3+1$ dimensional space.

\section{Non-Abelian Vortices and Topological $Z$-strings }
\label{vortices}

In this section, we discuss vortices in the first symmetry breaking in which only the triplets acquire VEVs.
In the first subsection, we discuss a non-Abelian vortex in the limit of the absence of the $U(1)_\Y$ gauge interaction.
In the second subsection, we discuss that non-Abelian vortices reduce to 
a $Z$-string or $W$-string when we turn on the $U(1)_\Y$ gauge interaction.

\subsection{Non-Abelian vortices in the absence of the $U(1)_\Y$ gauge interaction}
Here we assume $g_{\rm Y} =0$ at the starting and effect of $g_{\rm Y} \ne 0$ would be discussed in the next subsection. 
To construct vortices we only concentrate on the first phase transition as discussed in Eq.~(\ref{fullbreaking1}).
Since $\pi_1\l(G/H_3\r) = \mathbb{Z}_2$, we may have a vortex solution. In this section, to avoid complication, we set the doublet field zero. The consequences of interaction with doublet field will be discussed in the next section.

So we start with the Lagrangian density
 \begin{eqnarray}
 \label{Lvortex}
\mathcal{L} &=& - \frac{1}{8}\Tr W_{\mu\nu}^2 + \half \Tr(\D_\mu\Phi)^\dagger\D_\mu\Phi - V(\Phi) \\
V(\Phi) &=&\lambda_1\l( \Tr \Phi^\dagger\Phi - 3v_3^2\r)^2 + \lambda_2\l[3 \Tr \Phi^\dagger\Phi \Phi^\dagger\Phi - \Tr \l( \Phi^\dagger\Phi\r)^2\r]
\label{phipot}
\end{eqnarray}
where $W_{\mu\nu} = \p_\mu W_\nu - \p_\mu W_\mu - i g_{\rm W} [W_\mu, W_\nu] $  and  $ \D_\mu \Phi = (\p_\mu- i g_{\rm W} W_\mu^aT^a )\Phi$.   Here $g_{\rm W}$  is  the coupling constants of $SU(2)_\L$  gauge interactions.

For simplification we construct an infinitely long vortex along the $z$-axis with a cylindrical symmetry.  To derive a vortex solution let us start with the ansatz of $\Phi$  and $W_\mu$ as
\begin{eqnarray}
\label{vortexansatz}
\Phi_{\rm vortex} =  v_3 \left(
\begin{array}{ccc}
f(r) e^{i\theta}  &  0 & 0 \\
0  & g(r)  &0  \\
0  &0  &  f(r) e^{-i\theta} 
\end{array}
\right), \qquad W_i =  -\frac{\epsilon_{ij} x^j}{g_{\rm W} r^2}(1 + h(r))\left(
\begin{array}{ccccccc}
 1 & &  0 & & 0 \\
0  & &  0  & & 0  \\
0  & & 0  &  &  - 1
\end{array}
\right),
\end{eqnarray}
where $i=1, 2$ and $W_0$ and $W_3$ are taken to be zero.
Boundary conditions for profile functions are taken to be
\begin{eqnarray}
f(0) = 0,\,  f(\infty) = 1,\,  g'(0)=0,\,  g(\infty) = 1,\,  h(0) = -1,\,  h(\infty) = 0.
\end{eqnarray}
Here $(r, \theta)$ are the radius and azimuthal angle of the cylindrical coordinates. 
Let us first consider a large distance behavior of the vortex ansatz. From the above solution ansatz  we may write
\begin{eqnarray}
&&\Phi_{\rm vortex}(\theta, \infty) = v_3 \left(
\begin{array}{ccc}
 e^{i\theta}  &  0 & 0 \\
0  & 1  &0  \\
0  &0  &   e^{-i\theta} 
\end{array}
\right) = \Omega(\theta) \Phi_v, \quad 
\text{where} \,\, \Omega = e^{i\theta T_3}\in \frac{G}{H_3}.
\end{eqnarray}
 To find the behavior of the profile functions, 
 let us just put the above ansatz in the potential in Eq.~(\ref{phipot}) to yield
\begin{eqnarray}
V(\Phi) 
&=&  \lambda_1 v_3^4 \l( 2 f(r)^2  + g(r)^2 - 3\r)^2 + 2  \lambda_2 v_3^4\l[ f(r)^2 -   g(r)^2  \r]^2.
           \end{eqnarray}
           
The static Hamiltonian density
\begin{eqnarray}
\mathcal{H} &=& \int d^2x\l[\frac{1}{8}\Tr F_{ij}^2 + \half \Tr \l(D_i\Phi\r)^\dagger D_i\Phi + V(\Phi)\r]
\end{eqnarray}
can be expressed in terms of profile functions as
\begin{eqnarray}
\mathcal{H} &=&  2\pi \int r dr \l[ \half \frac{\l(\p_r h(r)\r)^2}{g_{\rm W}^2 r^2} +  v_3^2 \l\{ \l(\p_r f(r)\r)^2 + \frac{h(r)^2f(r)^2}{r^2}  + \half\l(\p_r g(r)\r)^2\r\}\r. \nonumber\\
&&\phantom{xxxxxxxxxxxxxx}\l. +  \lambda_1 v_3^4 \l[ 2 f(r)^2  + g(r)^2 - 3\r]^2 + 2  \lambda_2 v_3^4\l[ f(r)^2 -   g(r)^2  \r]^2\r].
\end{eqnarray}
This is actually the Hamiltonian density along the $z$-axis. Since all our fields are independent of $z$-coordinate we omit the $z$ integral. Let us rewrite the above Hamiltonian by defining $l =  \frac{\lambda_2}{\lambda_1}$, $\lambda_\rho = \frac{\lambda_1}{e^2}$ and $ \rh^2 = 2 e^2v_3^2 r^2$ as
\begin{eqnarray}
\label{energy_vportex1}
\mathcal{H}  &=&  2\pi  v_3^2 \,\,\times  \epsilon(\lambda_\rh, l), \nonumber \\ \epsilon(\lambda_\rh, l) &=&  \int \rh d\rh  \l[  \frac{\l(\p_\rh h(\rh)\r)^2}{\rh^2} +   \l\{  \l(\p_\rh f(\rh)\r)^2 +  \frac{h(\rh)^2f(\rh)^2}{\rh^2}  + \half\l(\p_\rh g(\rh)\r)^2\r\}\r. \nonumber\\
&&\phantom{xxxxxxxxxxxx}\l. + \frac{ \lambda_\rh}{2} \l[  \l\{ 2 f(r)^2  + g(r)^2 - 3\r\}^2 + 2  l \l[ f(r)^2 -   g(r)^2  \r]^2\r]\r].  
\end{eqnarray}
The equations of motion can be read off as
\begin{eqnarray}
&&- \rh \p_\rh \l[\frac{\p_\rh h(\rh)}{\rh}\r] +    f(\rh)^2h(\rh) = 0, \nonumber\\
&&-  \frac{1}{\rh}  \p_\rh \l[\rh \p_\rh f(\rh) \r]+  \frac{h(\rh)^2f(\rh)}{\rh^2}  + 2  \lambda_\rh  \l[ \l(2 + l \r) f(\rh)^2 + \l( 1 - l\r)  g(\rh)^2- 3\r]f(\rh)  = 0, \nonumber\\
&&-   \frac{1}{\rh}\p_\rh \l[\rh \p_\rh g(\rh) \r] + 2  \lambda_\rh \l[2 \l(1 - l\r) f(\rh)^2  +\l( 1 + 2 l \r) g(\rh)^2 - 3\r]g(\rh) = 0.
\end{eqnarray}
\begin{figure}[!htb]
\centering
\subfigure[\, ]{\includegraphics[totalheight=3.7cm]{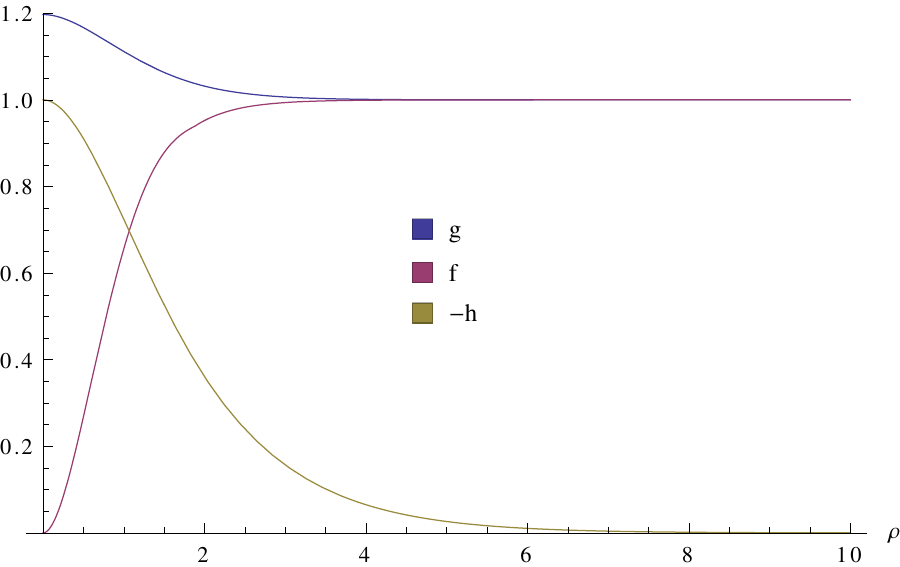}}
\subfigure[\, ]{\includegraphics[totalheight=3.7cm]{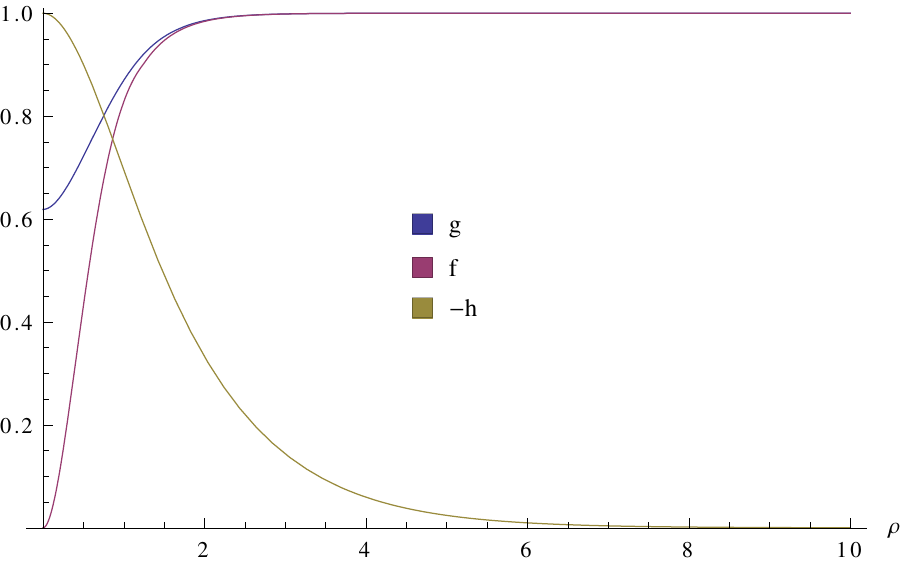}}
\subfigure[\, ]{\includegraphics[totalheight=3.7cm]{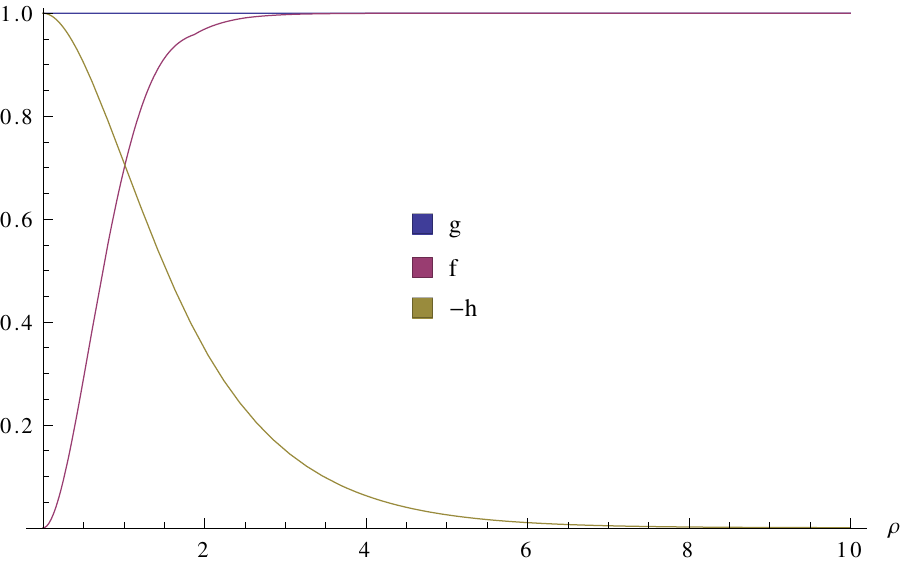}}
\subfigure[\, ]{\includegraphics[totalheight=3.7cm]{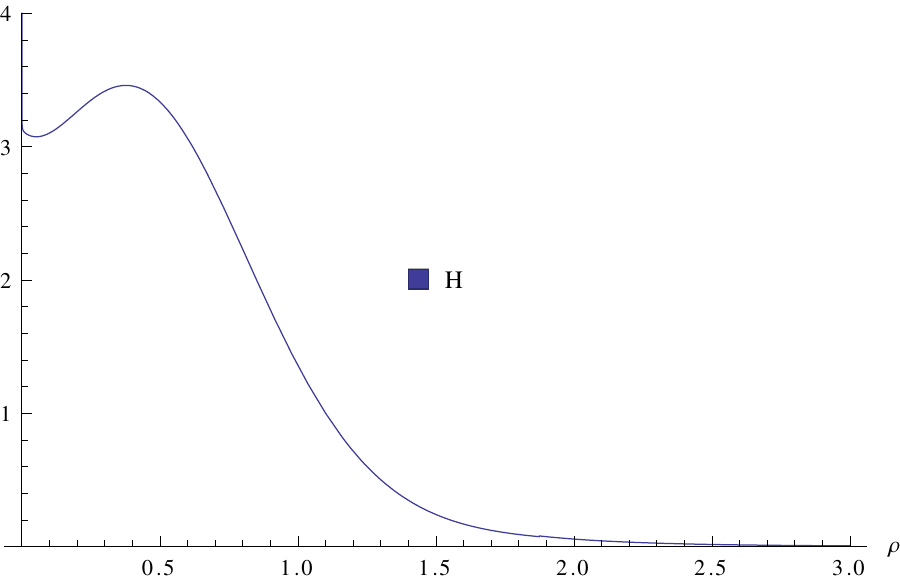}}
\caption{ This figure representing the plots of non-Abelian vortices discuss in section \ref{vortices}. In this computation we used 
$\lambda_1 = 0.2, g_{\rm W} = 0.63$. 
These are computed by  imposing the data  
$e  =  0.3 $, $\sin^2\theta_{\rm W} = 0.23$. 
We choose   $l = 0.5$ for figure (a),  $l = 5 $  for figure (b), 
$l = 1$ for figure (c). In figure (d) the energy density is plotted for $l = 5$ case.}
\label{Case1_f1f2}
\end{figure}

The tension of the vortex can be computed by inserting the profile function into the Eq.~(\ref{energy_vportex1}) and integrating over the $(x, y)$ plane. For the values of 
$\lambda_1 = 0.2, g_{\rm W} = 0.63$ 
we find 
$\epsilon(\lambda_\rh) \simeq 2 $.
 In Fig.~\ref{Case1_f1f2}, we show the profile functions for the cases of $l=0.5, 1, 5$ with fixing $\lambda_1=0.2, g_{\rm W}=0.63, e=0.3$, and ${\rm sin}^2\theta_{\rm W}=0.23$. 
 In the figure, we also plot the energy density as a function of $\rho$ for the case of $l=5$.


As we know, the VEV of $\Phi_v = v_3 \mathbf 1_{3\times 3}$ preserves the $SU(2)_\V$ custodial symmetry. However the vortex ansatz in Eq.~(\ref{vortexansatz})
breaks spontaneously the custodial $SU(2)_\V$ symmetry to a $U(1)$ subgroup inside the vortex due to the existence of two different profile behaviors, $f(r)$ and $g(r)$, inside the vortex core.
This generates infinite degenerate solutions which can be parametrize by an element on $S^2 \simeq SU(2)_\V/ U(1) $.
We may write a generic solution by a global  $SU(2)_\V$ transformation as
\begin{eqnarray}
\label{phin}
\Phi_{\rm vortex}(r, \theta, \xi^\a) &=& v_3 G(\xi^\a)\left(
\begin{array}{ccc}
 e^{i\theta} f(r)  &  0 & 0 \\
0  & g(r)  &0  \\
0  &0  &   e^{-i\theta} f(r)
\end{array}
\right) G^\dagger(\xi^a),\nonumber\\ W_i(r, \xi^\a) &=&  -\frac{\epsilon_{ij} x^j}{r^2}(1 + h(r)) \mathbf{\hat n}(\xi^\a), 
\end{eqnarray}
where  $ \mathbf{\hat n}(\zeta)$ is a unit vector oriented along a generic point on $SU(2)_\V/U(1) \simeq S^2$ defined  as 
\begin{eqnarray}
 \mathbf{\hat n}(\xi^\a) = G(\xi^\a)T^3 G^\dagger(\xi^\a) = \hat n^\a T^\a. 
\end{eqnarray}
Here $G(\xi^\a)$ is an element in the coset $S^2 \simeq SU(2)_\V/U(1)$, and $\xi^\a$ are the moduli 
parametrizing  $S^2$. 

These degenerated solutions can be varied slowly 
along $z$-axis with time with out changing the profile functions. So if we integrate the profile functions we land up with an effective action which is a nonlinear sigma model  defined   on 2D worldsheet (here $t$-$z$ plane) where 
fields are parametrized by the moduli parameters $\xi^\a$.

\subsection{Topological $Z$-strings and $W$-strings in the presence of the $U(1)_\Y$ gauge interaction}
In the above discussion, we have discussed the symmetries of potential, which is the same as to the case when there is no $B_\mu$ interaction or $g_{\rm Y} =0$. In this case all gauge fields are massive so 
we neglected their interactions. So we  find massless NG modes on  the vortex. However, in reality $g_{\rm Y} \ne 0$ and this breaks custodial symmetry explicitly 

the NG modes are lifted to become pseudo-NG modes, 
and consequently the most of non-Abelian vortices become unstable. 
The same phenomenon was first found for non-Abelian vortices in dense QCD \cite{Vinci:2012mc,Cipriani:2012hr} 
and later applied to supersymmetric QCD \cite{Konishi:2012eq}.

In this subsection we discuss the construction of vortices in the presence of the $B_\mu$ gauge field. 
The kinetic term of the scalar field is given as
$\D_\mu \Phi = (\p_\mu- i g_{\rm W} W_\mu^aT^a )\Phi + i g_{\rm Y} \Phi B_\mu T^3.$
So naturally there is a chance for the vortex flux to share the $B_\mu$ fields. As we know from  Eq.~(\ref{phin}) that the general vortex solution in the absence of $B_\mu$ field can be written 
with the flux directed along the unit vector $\hat n$ living on a sphere $S^2 \simeq SU(2)_\V/U(1)$. 
However, as it can be seen from the expression of covariant derivative in Eq.~(\ref{Dmu}) that the covariant derivative does not transform covariantly under the global transformation 
along generic direction on $S^2$ since the last term  breaks the degeneracy. 
However, there exist degenerate solutions on a subspace which consist of  the north and south pole
of the sphere. 
The equator circle also gives degenerate solutions, but they are energetically unstable different from the other two (north and south pole). We discuss the different configurations as follows. 
\subsubsection*{The $Z$-strings}
The $Z$-strings are defined on the north and south poles of the moduli space $S^2$. 
At the north pole the scalar field ansatz is given as
\begin{eqnarray}
\label{phiY}
\Phi_{\rm vortex} =  v_3 \left(
\begin{array}{ccc}
f(r) e^{i\theta}  &  0 & 0 \\
0  & g(r)  &0  \\
0  &0  &  f(r) e^{-i\theta} 
\end{array}
\right),
\end{eqnarray}
with the boundary condition$ f(0) = 0, f(\infty) = 1, g'(0)=0, g(\infty) = 1.$
Now by solving large distance condition $\D_i\Phi  \stackrel{r\rightarrow \infty}{\longrightarrow} 0.$
We may find the gauge field ansatz as
\begin{eqnarray}
\label{Zvortex}
&& {Z}_i  = \,\, \cos\theta_{\rm W} W_i^3 -  \sin\theta_{\rm W}   B_i   =  \,\,-\frac{\epsilon_{ij} x^j}{g_{\rm Z} r^2}(1 + h(r))\left(
\begin{array}{ccccccc}
 1  & &  0 & & 0 \\
0  & & 0  & & 0  \\
0  & & 0  & &   - 1
\end{array}
\right), \nonumber \\ 
&& h(0) = -1, \quad h(\infty) = 0, 
\end{eqnarray}
where $ \cos\theta_{\rm W} = \frac{g_{\rm W}}{\sqrt{g_{\rm W}^2+g_{\rm Y}^2}}$ and $ \sin\theta_{\rm W} = \frac{g_{\rm Y}}{\sqrt{g_{\rm W}^2+g_{\rm Y}^2}}$.  The difference with the Eq.~(\ref{vortexansatz}) in the expression of the ansatz
is that in Eq.~(\ref{Zvortex}) the coupling const $g_{\rm W}$ is replaced by $g_{\rm Z} = {\sqrt{g_{\rm W}^2+g_{\rm Y}^2}}$.

Similarly, at south pole the scalar field configuration would be given as
\begin{eqnarray}
\label{phiY}
\Phi_{\rm vortex} =  v_3 \left(
\begin{array}{ccc}
f(r) e^{-i\theta}  &  0 & 0 \\
0  & g(r)  &0  \\
0  &0  &  f(r) e^{i\theta} 
\end{array}
\right).
\end{eqnarray}
The gauge field solution would be same as Eq.~(\ref{Zvortex}) with negative sign in front. 

Since the vortex flux is completely determined by the flux of $Z_\mu$ field, these vortices can be called 
topological $Z$-strings.
The tension of the $Z$-string can be computed by inserting the profile function into the Eq.~(\ref{energy_vportex1}) with $g_{\rm W}$ is replaces by $g_{\rm Z}$ and integrating over the $(x, y)$ plane. For the values of $\lambda_1 = 0.2, g_{\rm Z} = 0.7283$ we find $E/z = 2\pi v_3^2 \times \,\, \epsilon\l(\frac{\lambda_1}{g_{\rm Z}^2}, l\r)$ and $ \epsilon\l(\frac{\lambda_1}{g_{\rm Z}^2}, l = 5\r) \simeq 2$.
\subsubsection*{The $W$-strings}
The $W$-strings are defined on the equator circle on $S^2$.
In this case, there is no flux sharing with $B_\mu$ field. 
The scalar field configurations are defined as
\begin{eqnarray}
\Phi_{W}(r, \theta, \xi_w) = G(\xi_w)\Phi(r, \theta, 0)  G^\dagger(\xi_w), \quad G(\xi_w) = \left(
\begin{array}{ccc}
 \frac{e^{-i\xi_w}}{\sqrt 2}  &  1 &  \frac{e^{-i\xi_w}}{\sqrt 2} \\
1  & 0  & - 1  \\
 \frac{e^{i\xi_w}}{\sqrt 2}  & -1  &   \frac{e^{i\xi_w}}{\sqrt 2}
\end{array}
\right) , \qquad 
\end{eqnarray}
where $\Phi(r, \theta, 0) = \text{diag}(f(r) e^{i\theta},  g(r), (r) e^{-i\theta} )$. The $W_\mu$ gauge field configurations are not diagonal in this case and is given as
\begin{eqnarray}
W_i =  -\frac{\epsilon_{ij} x^j}{g_{\rm W} r^2}(1 + h(r)) \left(
\begin{array}{cccccc}
0 & & \sqrt 2 e^{-i\xi_w} &  & 0  \\
\sqrt 2 e^{i\xi_w} &  & 0  & & \sqrt 2 e^{-i\xi_w}  \\
 0 &  & \sqrt 2 e^{i\xi_w} &  &  0
\end{array}
\right). \end{eqnarray}
Here $\xi_w$ is parametrizing the degenerate solutions along the equator circle, since $W$-string solutions are
constructed by global transformation of the solution written in Eq.~(\ref{vortexansatz}). The energy is the same as 
the vortex constructed in the case of $g_{\rm Y} = 0$. 
\begin{figure}[!htb]
\centering
\includegraphics[totalheight=6cm]{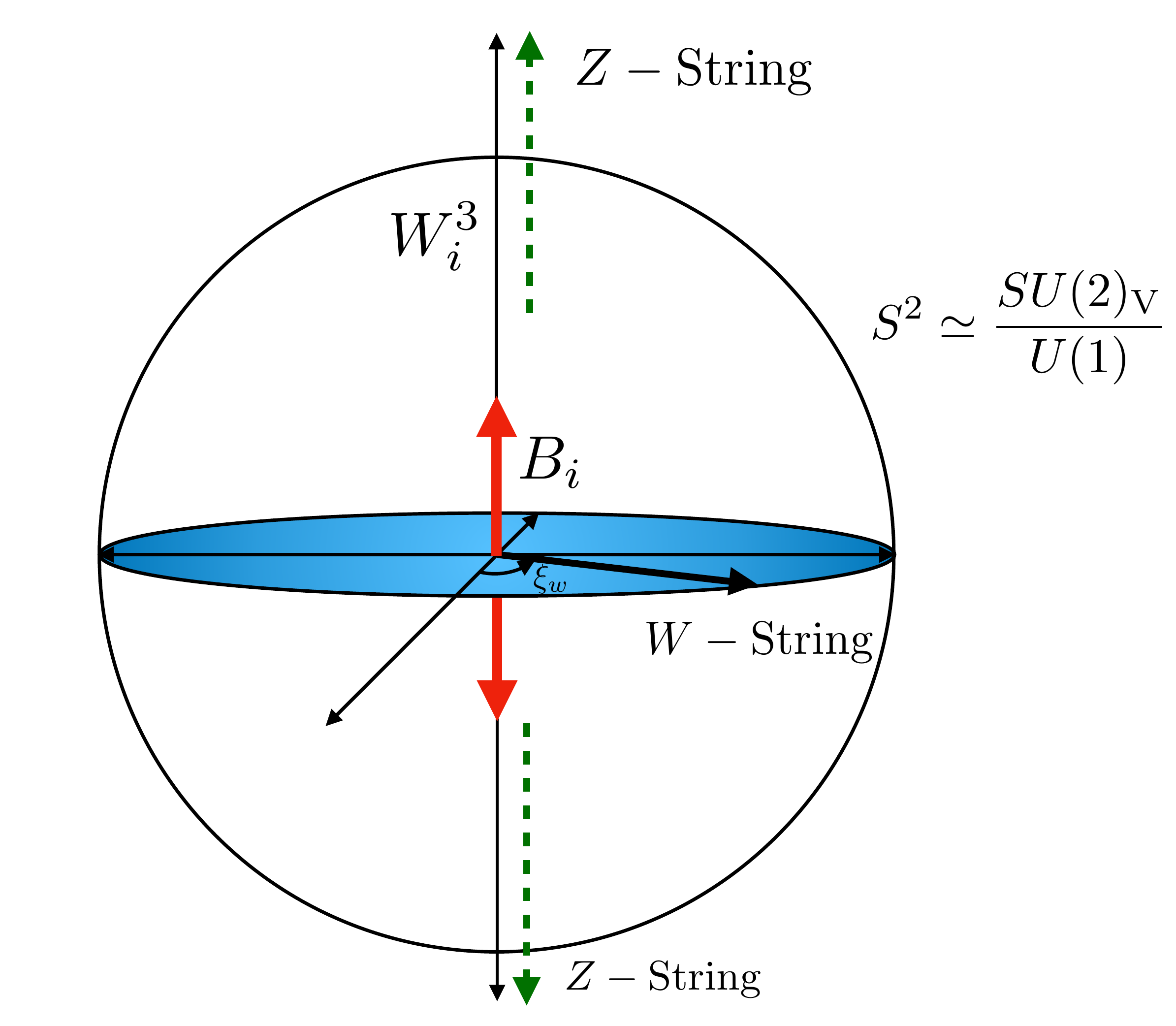}
\caption{A schematic view of $Z$- and $W$-strings on the moduli space $S^2$}
\label{WZstring}
\end{figure}
In Fig.~\ref{WZstring}, we show the schematic picture of $Z$- and $W$-strings on the moduli space of $S^2$.

\section{Non-Abelian Domain Walls Bounded by Non-Abelian Vortices}
\label{DVcomplex}

In Sec.~\ref{DWconstruction} we have constructed a stable domain wall in the GM model 
in the second symmetry breaking $H_3 \to H_2$ with a hierarchical symmetry breaking 
 $G \to H_3 \to H_2$.  
In the last section, we have constructed a vortex configuration in the first symmetry breaking $G \to H_3$, 
as described in Eq.~(\ref{fullbreaking1}).
In this section, we consider the most general case; 
the coexistence of domain walls and vortices in the full symmetry breaking. 
In the first subsection, we observe a behavior of a doublet around a vortex for preparation 
of the subsequent subsections. 
In the second subsection, we show that the general configuration 
is a partly global and partly local vortex in the vanishing limit of 
the interaction between the doublet and triplets.
In the third subsection, 
we show that it is a non-Abelian domain wall bounded by a non-Abelian vortex 
where $S^2$ moduli match at the junction line, 
if we turn off the $U(1)_{\rm Y}$ gauge coupling, 
while a $Z$-wall bounded by a $Z$-string.
Finally, in the fourth subsection, we calculate the quantum decay rate of a domain wall 
by quantum tunneling of creating a hole bounded by a closed vortex line. 

\subsection{Behaviour of a doublet encircling around a vortex}
\label{singlevaluedness}

We show in this subsection that when the doublet field acquires a VEV 
a pathology happens. 
To this end, 
we first investigate what happens when  
 a doublet field encircles around a vortex.
We note that the large distance behavior of a vortex configuration along 
the $z$-axis can be written as
\begin{eqnarray}
\Phi_{\rm vortex}(r=\infty,\theta) =  \Omega_3(\theta)\Phi_{\rm vortex}(r=\infty,\theta=0), \quad
\end{eqnarray}
where $\Omega_3$ is a holonomy acting on the triplet fields around a vortex, given by 
\begin{eqnarray}
\Omega_3(\theta) = P e^{i\int_0^\theta W\cdot dl}  = e^{i\theta T^3}. 
\end{eqnarray}
Here, we note that $\Omega_3$ is single-valued as usual $\Omega_3(2\pi)  = \mathbf{1}_{3\times 3}$.

On the other hand,
when a field $\eta(x)$ in a doublet representation encircles around a vortex, 
it receives a gauge transformation 
\begin{eqnarray}
\eta (r=\infty, \theta) = \Omega_2(\theta) \eta(r=\infty,\theta=0),\qquad  
\end{eqnarray}
where $\Omega_2$ is a holonomy acting on the doublet, given by
\begin{eqnarray}
\Omega_2(\theta) = P e^{i\int_0^\theta W\cdot dl}  =  e^{i\theta \tau^3}. 
\end{eqnarray}
In this case, it has a nontrivial holonomy when it encircles around a vortex 
\begin{eqnarray}
  \Omega(2\pi) = - {\bf 1}_{2\times 2} ,
\end{eqnarray}
and consequently 
\begin{eqnarray}
\eta (\theta = 2\pi) = -  \eta (\theta=0) . 
\end{eqnarray}
Therefore, the doublet field cannot become single-valued around a vortex, 
and so it has a nontrivial Aharanov-Bohm phase.

This brings us a pathology; the doublet field may not be allowed to acquire a VEV, and consequently 
the SM symmetry breaking could not occur in the presence of a vortex. 
This puzzle can be solved in two ways. 
One can make a vortex to a global vortex or one can create a domain wall.
The former happens when the interaction between doublet and triplet is negligible, 
namely when $\lambda_4$ is small in Eq.~(\ref{potential}).
The latter happens when the interaction term with $\lambda_4$ is relevant.
In the following subsections, we discuss these two cases separately.

\subsection{A composite of global-local vortex configuration}
\label{global}

Here we discuss a vortex configuration which develops after the second symmetry breaking in a special circumstance. 
We switch off the interaction term between the doublet and triplets, {\it i.e.}, $\lambda_4 = 0$ and also hypercharge $ g_{\rm Y} =0$. In this case, the triplets and doublet fields interact via the $SU(2)_\L$ gauge interaction only. Since in the absence of the $\lambda_4 $ term, the right actions on the doublet and triplets are independent, which we denote
$SU(2)_{\R_1}$ and $SU(2)_{\R_2}$, respectively. 
So in this case, we start with full symmetry breaking group as
$G(R_1, R_2) = SU(2)_\L \times SU(2)_{\R_1} \times SU(2)_{\R_2}$.
Now we set our vacuum expectation values of the  fields as before as $\Phi_v = v_3 \mathbf 1_{3\times 3}$ and $\Psi_v = v_2 \mathbf 1_{2\times 2}$, this breaks $G(R_1, R_2)$ to the diagonal group
$ H_2(R_1, R_2) = \mathbb{Z}_2\times SU(2)_{\L + \R_1+\R_2}$.
 The full symmetry breaking is discussed in Appendix \ref{gvsymmetry} in details.  
The existence of a new vortex solution can be understood, 
if we note that 
the vacuum manifold
 \begin{eqnarray}
    \frac{G(R_1, R_2)}{H_2(R_1, R_2)} 
  = \frac{SU(2)_\L \times SU(2)_{\R_1} \times SU(2)_{\R_2}}{\mathbb{Z}_2\times SU(2)_{\L + \R_1 + \R_2}}
\end{eqnarray}
allows the first homotopy group 
\begin{eqnarray}
 \pi_1\l(\frac{G(R_1, R_2)}{H_2(R_1, R_2)}\r) 
 = \mathbb{Z}_2. 
\end{eqnarray}
The $ \mathbb{Z}_2$ factor of $H_2$ contains the element $(-1, 1, -1)$ which would be responsible for our new vortex solution and this would keep the doublet field single valued. This is because rotation around existing flux tube
generate a negative sign where the existence of a global rotation in $SU(2)_{\R_2}$ would generate other negative sign to compensate the other.  
In order to construct a vortex solution, 
let us define our doublet ansatz as
\begin{eqnarray}
\Psi_{\rm vortex}(r, \theta) = v_2 \psi(r) \exp (i \theta \sigma^3), \quad 
\end{eqnarray}
with the boundary conditions for the profile function $\psi$, given by
\begin{eqnarray}
\psi(0) = 0, \quad \psi(\infty) = 1.
\end{eqnarray}
With this ansatz for the doublet field, 
we use the ansatz for the triplet fields and gauge field given in Eq.~(\ref{vortexansatz}). 
The large distance behavior  of the doublet field can be expressed as
\begin{eqnarray}
\label{psiansatz}
\Psi_{\rm vortex}(\infty, \theta) = Pe^{i\int^\theta_0 W\cdot dl} \Psi_v e^{i \theta \frac{\sigma^3}{2} }. 
\end{eqnarray}
From this expression it is now clear that the full loop of $\pi_1(G/H_2)$ has two contributions. 
One from an $SU(2)_\L$ gauge transformation accompanied with a gauge flux 
and the other is a global transformation of $SU(2)_{\R_2}$. 
Therefore, our vortex is a half local and half global vortex with a magnetic flux.

\begin{figure}[!htb]
\centering
\subfigure[\, ]{\includegraphics[totalheight=3.5cm]{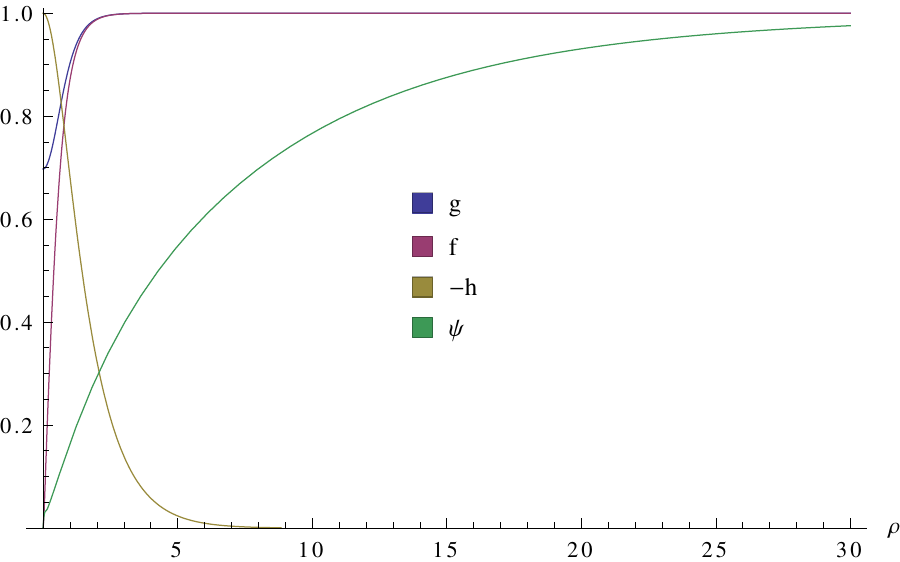}}
\subfigure[\, ]{\includegraphics[totalheight=3.5cm]{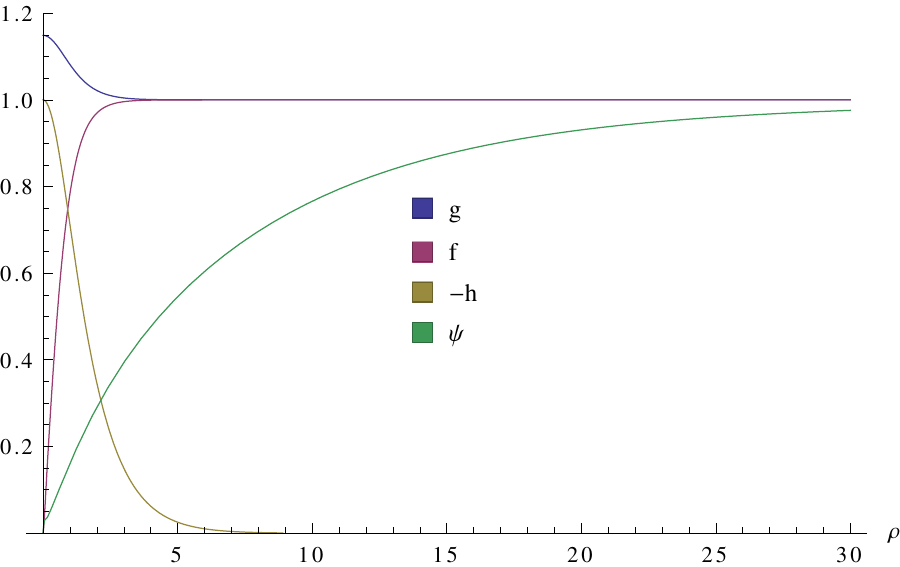}}
\subfigure[\, ]{\includegraphics[totalheight=3.5cm]{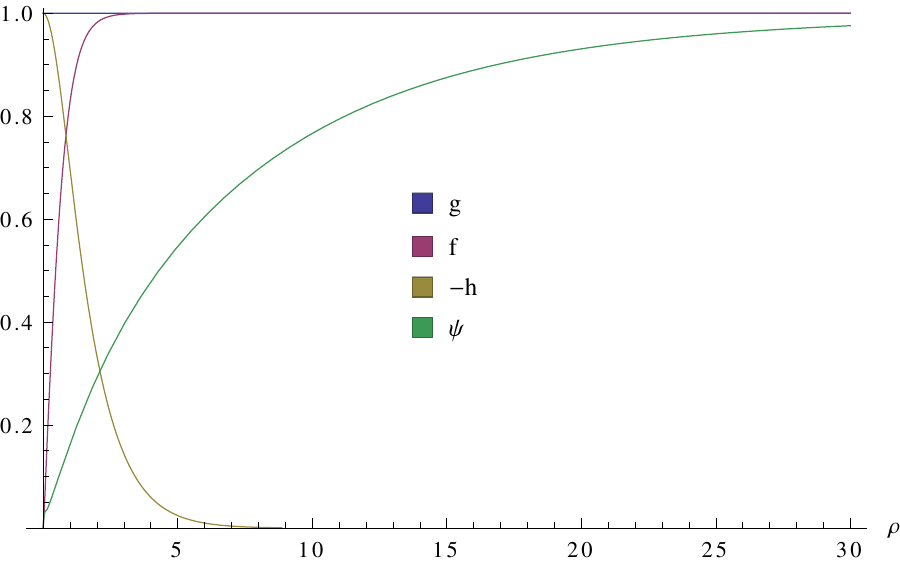}}
\subfigure[\, ]{\includegraphics[totalheight=3.5cm]{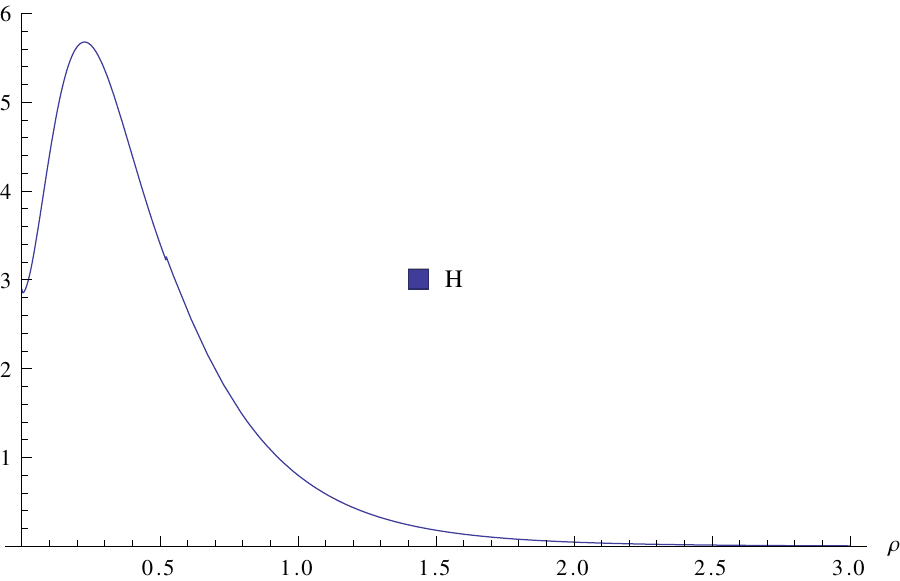}}
\caption{The plots of the profile functions of a global vortex described in Subsec.~\ref{global}. 
In this calculation, we have used $\lambda_1 = 0.2, \lambda_3 = 0.2, g_{\rm W} = 0.63$. These are computed by  imposing the data 
 $e  =  0.3 $, $\sin^2\theta_{\rm W} = 0.23$. We choose   $l = 5$ for the panel (a),  $l = 0.5 $  for the panel (b), 
$l = 1$ for the panel (c). In the panel (d), the energy density is plotted for $l = 5$.}
\label{psistring}
\end{figure}

By using the vortex aznsatz in Eqs.~(\ref{vortexansatz}) and (\ref{psiansatz}), 
we may rewrite the  Hamiltonian, by defining 
$l =  \frac{\lambda_2}{\lambda_1}$, $\lambda_\rho = \frac{\lambda_1}{e^2}$ , $\tilde\lambda_3 = \frac{\lambda_3}{e^2}$and $ \rh^2 = 2 e^2v_3^2 r^2$ , $ v = \frac{v_2}{v_3}$, as
\begin{eqnarray}
\mathcal{H}  &=&  2\pi  v_3^2 \int \rh d\rh  \l[  \frac{\l(\p_\rh h(\rh)\r)^2}{\rh^2} +   \l\{  \l(\p_\rh f(\rh)\r)^2 +  \frac{h^2f(\rh)^2}{\rh^2}  + \half\l(\p_\rh g(\rh)\r)^2\r\}\r.\nonumber \\
&&\l. + \frac{ \lambda_\rh}{2} \l[  \l\{ 2 f(r)^2  + g(r)^2 - 3\r\}^2 + 2  l \l[ f(r)^2 -   g(r)^2  \r]^2\r]\r] \nonumber \\ && + \pi  v_2^2 \int \rh d\rh  \l[  \l(\p_\rh \psi(\rh)\r)^2 +  \frac{(1 - h)^2\psi(\rh)^2}{4\rh^2}  
 + \tilde\lambda_3 v^2   \l[ \psi^2  -1\r]^2 \r]. 
\end{eqnarray} 
The equations of motion can be read off as
\begin{eqnarray}
&&- \rh \p_\rh \l[\frac{\p_\rh h(\rh)}{\rh}\r] +    f(\rh)^2h(\rh) + \frac{v^2}{8} \l(h(\rh)-1\r)\psi(\rh)^2= 0, \nonumber\\
&&-  \frac{1}{\rh}  \p_\rh \l[\rh \p_\rh f(\rh)\r] +  \frac{h(\rh)^2f(\rh)}{\rh^2}  + 2  \lambda_\rh  \l[ \l(2 + l \r) f(\rh)^2 + \l( 1 - l\r)  g(\rh)^2- 3\r]f(\rh)  = 0, \nonumber\\
&&-   \frac{1}{\rh}\p_\rh \l[\rh \p_\rh g(\rh) \r] + 2  \lambda_\rh \l[2 \l(1 - l\r) f(\rh)^2  +\l( 1 + 2 l \r) g(\rh)^2 - 3\r]g(\rh) = 0, \nonumber\\
&&-  \frac{1}{\rh}  \p_\rh \l[\rh \p_\rh \psi(\rh)\r] +  \frac{(1 - h(\rh))^2\psi(\rh)}{4\rh^2}  + 2 \tilde\lambda_3 v^2   \l[ \psi^2  -1\r]\psi = 0
\end{eqnarray}
A numerical solution is shown in Fig.~\ref{psistring}.

\subsection{Domain wall bounded by a vortex}

In the last subsection we have discussed the case in which the $\lambda_4$ interaction term 
between the doublet and triplets is negligible so that 
the global symmetries acting on  the doublet and triplets become independent 
and so the global symmetry is enhanced. 
In this subsection we consider the case in which the $\lambda_4$ term is relevant 
so that  the global symmetries acting on the doublet and triplets are locked: 
$SU(2)_{\R_1} = SU(2)_{\R_2}$. 
We show that in this case the global vortex in the last subsection transforms to be
a vortex-domain wall composite. 
We start with $g_{\rm Y} =0$  just to understand the orientational zero modes.
The effect of hypercharge, i.e.  $g_{\rm Y} \ne 0$  would be discussed later. 

\begin{figure}[!htb]
\centering
\subfigure[\, ]{\includegraphics[totalheight=3.5cm]{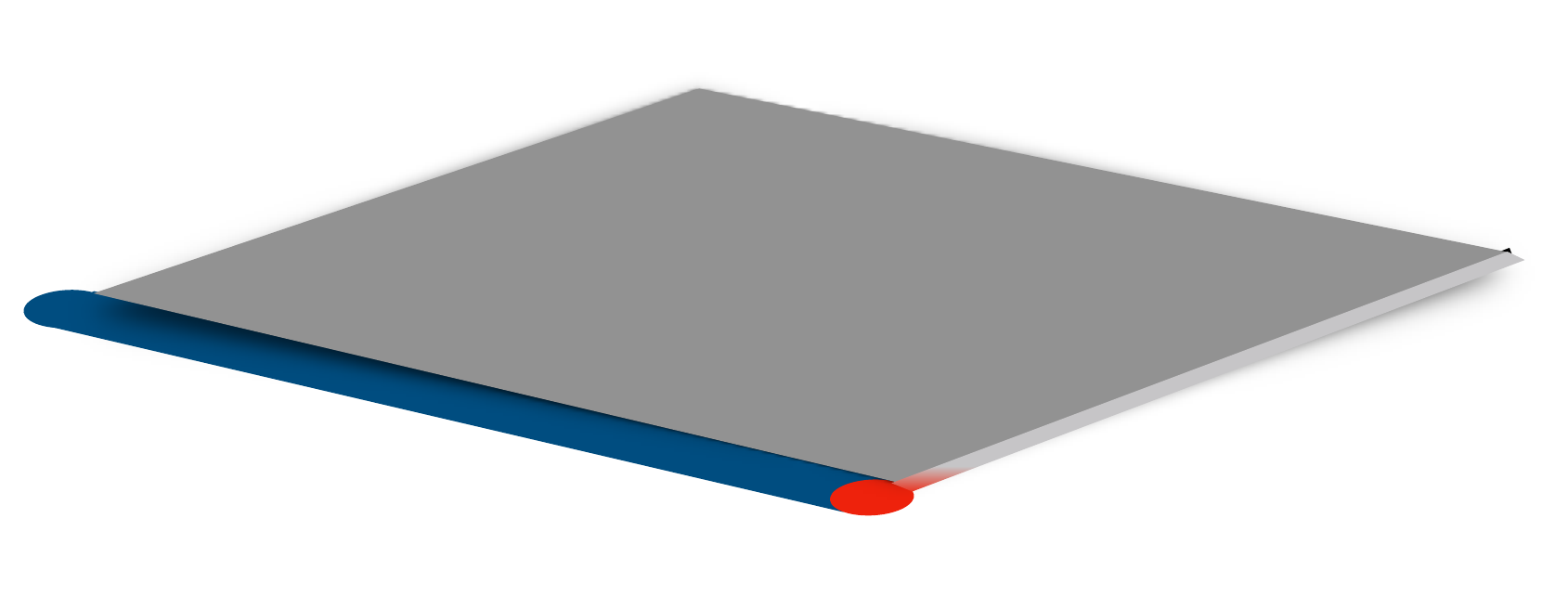}}
\label{dwvort1}
\subfigure[\, ]{\includegraphics[totalheight=3.5cm]{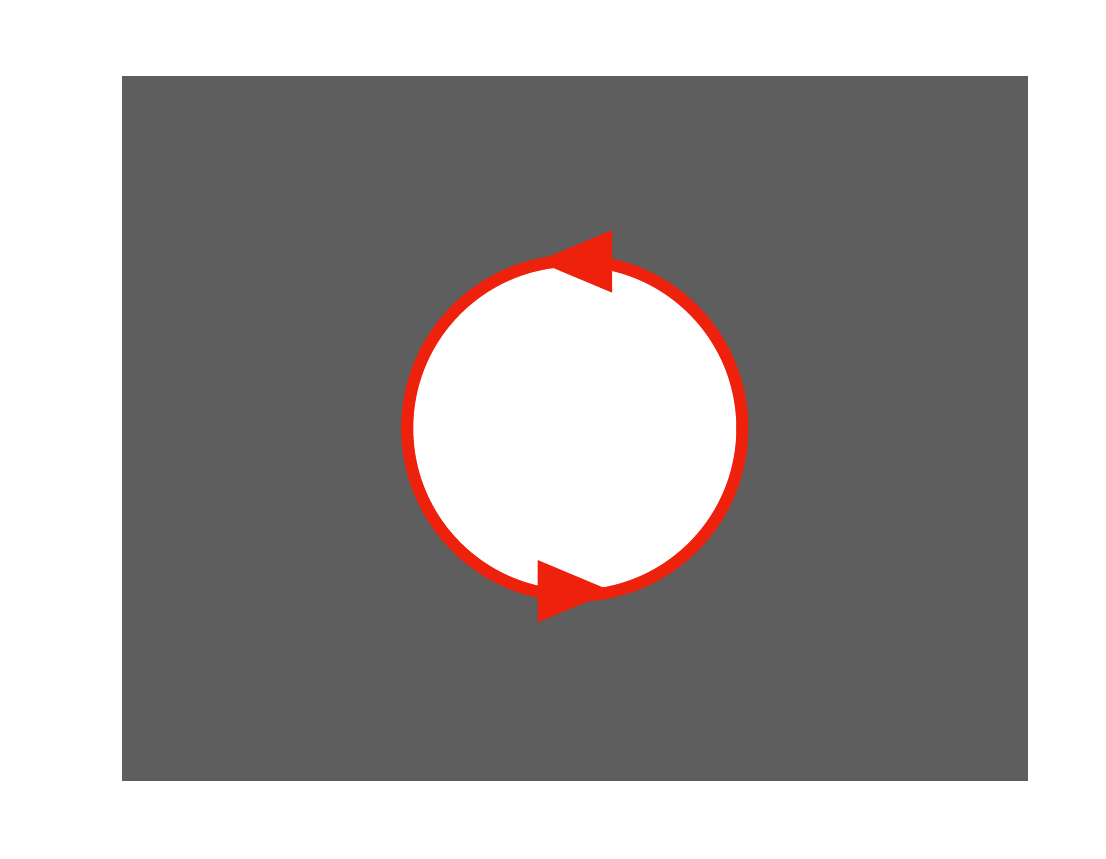}}
\label{dwvort2}
\caption{ The schematic diagrams of (a) Domain wall bounded by flux tube (b) Flux tube hole creation in domain wall
\label{fig:wall-vortex}
}
\end{figure}

In Subsec. \ref{singlevaluedness} we have discussed the puzzle  
that VEV of $\Psi$ might become multivalued around the vortex. 
Now we shall show that this problem can be cured by a creation of a domain wall in the doublet 
so that the total configuration becomes a vortex attached by a domain wall. 
As we discussed in Sec.~\ref{DWconstruction}, at the second symmetry breaking $H_3 = \mathbb{Z}_2 \times SU(2)_\V$ is broken down to $SU(2)_\V$. This gives $\pi_0(H_3/H_2) = \mathbb{Z}_2$, which confirms the existence of domain wall solution.
To find a domain wall attached to the vortex, we start with the ansatz
\begin{eqnarray}
\label{dw-vortex}
\Psi(x)_{{\rm dw}-{\rm vortex}} = \frac{v_2}{\sqrt 2}\psi(r) 
\left(
\begin{array}{ccc}
  e^{i \l[\frac{\phi(\theta) + \xi(\theta)}{2}\r]} &   0   \\
  0 &   e^{- i \l[\frac{\phi(\theta) + \xi(\theta)}{2}\r] }
\end{array}
\right)
\end{eqnarray}
where $\xi(\theta)$ and $\phi(\theta)$, 
both of which change from $0$ to $2\pi$ when one goes around a vortex, 
are contributions from the gauge transformation 
and the global $SU(2)_{\rm R}$ transformation, respectively. 
In this subsection, we study a large distance behavior of the system. 
The vortex solution at large distance behaves as
\begin{eqnarray}
\label{vortex-dw}
\Phi_{\rm vortex} \sim  v_3 \exp[i\xi(\theta)T^3], \qquad 
 W_i \sim \frac{\p_i\xi(\theta)}{g_{\rm W} }T^3.
\end{eqnarray}
We insert  the field configurations in Eqs.~(\ref{dw-vortex}) and (\ref{vortex-dw}) into 
each term of the potential term in Eq.~(\ref{potential}):
\begin{eqnarray}
V(\Phi,\Psi) &= &  \lambda_3\l( \Tr \Psi^\dagger\Psi - v_2^2\r)^2
+ \lambda_4\l( \Tr \Psi^\dagger\Psi \Tr \Phi^\dagger\Phi - 2 \Tr( \Psi^\dagger \tau^a \Psi \tau^b)  \Tr (\Phi^\dagger T^a \Phi T^b)\r) \qquad
\end{eqnarray} 
to yield
\begin{eqnarray}
 \Tr\l( \Psi^\dagger \tau^a \Psi \tau^b\r) =   \frac{v_2^2}{4} \psi^2  {R(\tau)}_{ab}, &&  \quad R(\tau) =  \left(
\begin{array}{ccc}
  \cos(\phi + \xi)  & -\sin(\phi + \xi) & 0 \\
 \sin(\phi + \xi)   &  \cos(\phi + \xi) & 0  \\
0  & 0  &  1
\end{array}
\right),  \\
\Tr \l(\Phi^\dagger T^a \Phi T^b\r) = 2v_3^2 R(T)_{ab} , && \quad R(T) = \left(
\begin{array}{ccc}
  \cos\xi   & - \sin\xi & 0 \\
 \sin\xi   &  \cos\xi  & 0  \\
0  & 0  &  1
\end{array}
\right) . 
\end{eqnarray}
We thus obtain
\begin{eqnarray}
\label{DWvortpot}
V(\psi,\phi) = v_2^2  \l[ \lambda_3 v_2^2 \l(  \psi(\theta)^2 - 1\r)^2 + \mu^2 \psi^2\l( 1 - \cos \phi(\theta)\r)  \r], 
\end{eqnarray}
where we have defined $\mu^2 = 2 \lambda_4 v_3^2$ as before in Eq.~(\ref{DWH}).  This potential is the same as what was found in Eq.~(\ref{DWH}), 
although the argument here is the angle $\theta$ around the vortex 
while it was one spatial direction $x$ in Eq.~(\ref{DWH}).
The ground state is given by $\psi = 1, \phi = 2 n\pi$.  As we have found in Sec.~\ref{DWconstruction} that this potential gives a sine-Gordon domain wall solution.   
This domain wall is attached to the infinitely long vortex along the $z$-axis centered at the origin, 
as schematically shown in 
Fig.~\ref{fig:wall-vortex} (a). 
When encircling around the vortex, the doublet field changes sign while passing through the domain wall  placed at $\theta = \theta_c$.
Therefore, the existence of domain wall solves the problem that the doublet might become multi-valued.

In our hierarchy symmetry breaking, 
the width of the vortex $\delta_{\rm vortex} \sim (v_3\sqrt\lambda_1)^{-1}$ is much smaller than
the wall width $\delta_{\rm dw} \sim \mu^{-1}$.

We have shown that
domain walls are bounded by $\mathbb{Z}_2$-strings described in Sec.~\ref{vortices}. One thing one should point out here that  our strings are not global strings but are flux-tubes.
This construction is in contrast to axionic string-domain wall composite,
in which strings are global strings.

Let us discuss here the orientational zero modes and effect of hypercharge, {\it i.e.}, $g_{\rm Y}\ne 0$. The computations done above do not depend on the orientation of vortex and domain wall. They can be oriented together in generic directions on 
$S^2 \simeq SU(2)_\V/U(1)$ 
since the $\lambda_4$ term keeps the triplet and the doublet in the same direction.  Oscillations of the zero modes would suppose to flow from the vortex worldsheet  to the domain wall surface and vice  versa. Actual dynamics would be described by effective action of the NG modes,
which will be the $O(3)$ model with a boundary.
 
 When $g_{\rm Y}\ne 0$ the NG modes are lifted. So at lowest energy, we have only a $Z$-string-domain wall.

\subsection{Quantum decay of a domain wall}

The domain wall construction discussed in this paper will not be stable after full symmetry breaking 
$\pi_0(G/H_2) = 0$. So according to Kibble \cite{Kibble:1982dd}, a hole would be created locally due to
 local  thermal or quantum fluctuations.
This hole is bounded by a closed vortex string as schematically shown in Fig.~\ref{fig:wall-vortex} (b). 
A hole smaller than some critical scale $R_c$ will be destroyed. 
However, there can be some hole creation with length scale more than 
$R_c$. Then this hole would start growing and the wall would become unstable. 
The decay rate can be computed using method described in Refs.~\cite{Kibble:1982dd, Vilenkin:2000jqa}.  The decay probability at zero temperature  is given as 
$
\Gamma \sim  e^{-S}
$ where
$S$ is the Euclidean action of the bounce solution  corresponding to tunneling process. In this case $S$ is given as
\begin{eqnarray}
S = 4\pi R^2 T_{\rm vortex} - \frac{4}{3} \pi R^3 T_{\rm dw}.
\end{eqnarray}
Here $T_{\rm vortex} = 2 \pi v_3^2\epsilon$ $(\epsilon \sim 2)$ is the vortex energy per unit length as defined in Eq.~(\ref{energy_vportex1}) and $T_{\rm dw} =  4\mu v_2^2$ is 
the energy of the domain wall per unit area, defined in Eq.~(\ref{DWH}). So $R_c =  \frac{2 T_{\rm vorex}}{T_{\rm dw}}$, and 
\begin{eqnarray}
S_c &&=  \frac{16\pi}{3} \l(\frac{ T_{\rm vortex}^3}{T_{\rm dw}^2}\r) = \frac{4 \pi^4}{3  \lambda_4}  \l(\frac{  v_3}{  v_2}\r)^4 \epsilon^3. 
\end{eqnarray}
In the case of $v_3 >> v_2$ and $\lambda_4 < 1$ , $\Gamma$ must be a  small number. 
So domain walls can be locally stable at zero temperature. However, this is a minimal estimation of the decay rate because this computation is valid when $S$ is very large so that other interactions can be neglected \cite{Weinberg:1996kr}. So when $v_3$ is close to $v_2$ this analysis may not be very practical.  We also did not take into account the finite temperature effect.
Due to interaction with other fields and at finite temperature this probability will be different and would be discussed elsewhere.

\section{Summary and Discussions}
\label{summary}
In this paper, we have discussed topological defects in the GM model. This model contains three Higgs triplets in addition to the usual Higgs doublet. 
We studied the spontaneous
breaking of $G=SU(2)_\L \times SU(2)_\R$ in two stages, 
 namely a hierarchical symmetry breaking 
$G \to H_3 \to H_2$ first triggered by the triplets followed by the doublet 
subsequently. 
The order parameter manifold
has nontrivial homotopy groups  
$\pi_1(G/H_3) =  \mathbb{Z}_2$ and  
$\pi_0(H_3/H_2) =  \mathbb{Z}_2$ supporting a ${\mathbb Z}_2$ vortex and a domain wall, respectively.  
We have first solved the vortex profile functions numerically
for the case of an axially symmetric infinitely long vortex in the decoupling limit of 
the $U(1)_\Y$ gauge field. 
In this case, the custodial $SU(2)_\V$ is spontaneously broken inside the vortex core
generating the $S^2$ NG modes localized around the vortex core. 
These modes correspond to a non-Abelian magnetic flux confined inside the vortex core.
When the $U(1)_\Y$ gauge coupling is taken into account, 
the $SU(2)_\V$ custodial symmetry is explicitly broken 
and  
the $S^2$ moduli space is lifted, 
leaving a stable $Z$-string and an unstable $W$-string as solutions.
All vortices including $W$-strings fall into a topologically stable $Z$-string, 
in contrast to the SM in which $Z$-strings are non-topological and are unstable in the realistic parameter region.
We then have discussed the vortex-domain wall complex,
in which the $S^2$ moduli of both the vortex and domain wall match at the junction line.
The vortex stable in the first symmetry breaking is attached by a 
 domain wall appearing in the second symmetry breaking, and 
consequently domain walls can decay through quantum tunneling by 
creating a hole bounded by a closed vortex line.  
We have calculated the decay rate at zero temperature which is found to be small.

Several discussions are addressed here.
In this paper, we have discussed only the Higgs sector. 
If we include the fermion sector, there are several interesting physics.
First, vortices \cite{Jackiw:1981ee} 
and domain walls \cite{Jackiw:1975fn} would have fermion zero modes around their cores, 
as an electroweak $Z$-string in the SM 
\cite{Vachaspati:1992mk,Moreno:1994bk,Earnshaw:1994jj,Garriga:1994wb,Naculich:1995cb,Liu:1995at,Starkman:2000bq,Starkman:2001tc,Graham:2011fw}
and a non-Abelian vortex in dense QCD 
\cite{Yasui:2010yw,Fujiwara:2011za}. 
For the former, it was argued that fermion zero modes may distabilize $Z$-strings, 
but in our case strings are stable (at the first symmetry breaking) because they are topological.
For the latter, these fermion zero modes would interact with NG modes \cite{Chatterjee:2016ykq}, 
and so a similar would happen in our case.

Second, fermions scattering off a non-Abelian vortex 
may receive an Aharanov-Bohm phase, 
as the cases of 
an electroweak $Z$-string \cite{Lo:1994fm,Nagasawa:1996sg}
 and 
a non-Abelian vortex in dense QCD \cite{Chatterjee:2015lbf} 
(see also Ref.~\cite{Evslin:2013wka} for the same situation in supersymmetric QCD).
All together, the exchange of multiple vortices with fermion zero modes may have nontrivial non-Abelian statistics as the case of dense QCD 
  \cite{Yasui:2010yh,Hirono:2012ad,Yasui:2011gk,Yasui:2012zb}.

The interaction of the electromagnetic waves and topological defects found in this paper may be important for a possibility of searches for these objects, such as cosmic microwave backgrounds. The interaction of the electromagnetic waves and a non-Abelian vortex through charged zero modes localized around the vortex was studied in dense QCD, in which case a vortex lattice is shown to behave as a polarizer \cite{Hirono:2012ki}. 
In our case, the interaction with a domain wall through the charged zero modes localized around it must be the most important possibility.
 The $Z$-string (or $Z$-string-domain wall composite) does not interact with electromagnetic gauge field $A_\mu$ at low energies.   Since NG modes are massive the high energy electromagnetic waves can excite NG modes  which interacts with $A_\mu$.  So in this sense the  $Z$-string(-domain wall composite) could be considered as a `gray-matter' element.

Electroweak baryogenesis by electroweak strings  
\cite{Brandenberger:1992ys} does not work in the SM
\cite{Nagasawa:1996gy}.
This problem may be rescued  
since our 
$Z$-strings are topologically stable in the first symmetry breaking.
If there is a long enough period between the first and second symmetry breakings
there could be enough baryogenesis.

Gravitational waves from decay of domain walls and vortices will give important signature of this scenario~\cite{Saikawa:2017hiv}.
Depending on the tension of domain walls and the temperature when they annihilate, amount of gravitational waves emitted from those could be significant 
enough to be detected by ongoing and future experimental searches. Detailed study of the spectrum of gravitational waves in the GM model will be studied elsewhere.

Other than vortices and domain walls, there may exist stable 
monopoles (instantons) and Skyrmions as composite states, in contrast to the SM.
Stable monopoles may exist as a kink on a $Z$-string since 
$S^2$ moduli are lifted 
due to the $U(1)_{\rm Y}$ gauge coupling,
leaving two points, the north and south poles, 
corresponding two $Z$-strings.
In this case, a monopole is attached by two $Z$-strings on both sides 
and so is stable.
Instantons may exist as lumps inside a vortex \cite{Eto:2004rz}, 
but they may be unstable because of the potential induced from the $U(1)_{\rm Y}$ gauge coupling.
On the other hand, Skyrmions may exist as lumps inside a domain wall 
(called domain wall Skyrmions)
\cite{Nitta:2012wi,Nitta:2012rq,Gudnason:2014nba} (see also \cite{Eto:2015uqa}) 
whose effective theory should be an $O(3)$ sigma model.

\appendix
\section{Symmetry of Global Vortices}
\label{gvsymmetry}
Here we discuss the symmetry and symmetry breaking in the absence of the interaction between the doublet and triplets, {\it i.e.}, $\lambda_4 = 0$ and  $ g_{\rm Y} =0$. In this case the the triplets and doublet fields interact only through the $SU(2)_\L$ gauge interaction. Since in this case the right symmetries act independently on
the triplets and doublet. 
We denote the right action groups on the triplets and doublet by 
$SU(2)_{\R_1}$ and $SU(2)_{\R_2}$, respectively.  
So in this case the full symmetry group is given as
\begin{eqnarray}
G(R_1, R_2) = SU(2)_\L \times SU(2)_{\R_1} \times SU(2)_{\R_2}. 
\end{eqnarray}
Now the VEV of the triplet fields $\Phi_v = v_3 \mathbf 1_{3\times 3}$ 
breaks $G(R_1, R_2)$ to (see the footnote $^{\ref{center}}$) 
\begin{eqnarray}
&& H_3(R_1, R_2) = (\mathbb{Z}_2)_{\rm -L+R_1} \times SU(2)_{\L + \R_1} \times SU(2)_{\R_2}, \nonumber \\ 
&& \quad  (\mathbb{Z}_2)_{\rm -L+R_1} =  \l\{(1, 1,1), (-1, 1,1)\r\}.
\end{eqnarray}
The first two entries in the elements of $ \mathbb{Z}_2$ is arising from the center of 
$SU(2)_\L$ and $SU(2)_{\rm R_1}$.
This $ \mathbb{Z}_2$ is unbroken because for the center is identified in the triplet representation. 

 Now let us discuss the second symmetry breaking by 
 $\Psi_v = v_2  \mathbf 1_{2\times 2}$, 
 which breaks $H_3$ further into 
\begin{eqnarray}
\label{H2R2}
&& H_2 = (\mathbb{Z}_2)_{\rm -L+R_1-R_2} \times SU(2)_{\L + \R_1 + \R_2}, \nonumber\\
\quad 
&& \quad (\mathbb{Z}_2)_{\rm -L+R_1-R_2} = \l\{(1, 1, 1), (-1, 1, -1)\r\}, 
\end{eqnarray}
Note that $(\mathbb{Z}_2)_{\rm -L+R_1-R_2}$ is different from  $(\mathbb{Z}_2))_{\rm -L+R_1}$ in $H_3$. 

To understand this let us write down the unbroken elements of the full center of $G(R_1, R_2)$, which is $ \mathbb{Z}_2\times \mathbb{Z}_2\times \mathbb{Z}_2$, as
\begin{eqnarray}
\l\{(1, 1, 1),(-1, -1, -1),(1, -1, 1),(-1, 1, -1)\r\}. 
\end{eqnarray}
All the nontrivial elements of this group have order two. 
We may rewrite these elements as an internal direct product of two $\mathbb{Z}_2$ subgroups as
\begin{eqnarray}
 \mathbb{Z}_2\times \mathbb{Z}_2 = \l\{(1, 1, 1), (-1, -1, -1)\r\} \times \l\{(1, 1, 1), (-1, 1, -1)\r\} . 
\end{eqnarray}
 This is an internal direct product in which the identity element is shared. 
 The first $ \mathbb{Z}_2$ factor is the center of the diagonal subgroup $SU(2)_{\L + \R_1 + \R_2}$, 
 while the second factor 
is $(\mathbb{Z}_2)_{\rm -L+R_1-R_2} $ in Eq.~(\ref{H2R2}). 

The existence of the new vortex solution can be understood if we see the fundamental group of the full symmetry breaking as
 \begin{eqnarray}
\pi_1\l(\frac{G(R_1, R_2)}{H_2(R_1, R_2)}\r) = 
\pi_1\l(
\frac{SU(2)_\L \times SU(2)_{\R_1} \times SU(2)_{\R_2}}{\mathbb{Z}_2\times SU(2)_{\L + \R_1 + \R_2}} 
\r) 
= \mathbb{Z}_2. 
\end{eqnarray}

\end{document}